\documentclass{aa}  
\usepackage{txfonts}
\usepackage{graphicx}	
\usepackage{newtxtext}

\usepackage[colorlinks,citecolor=blue, menucolor=blue, urlcolor=blue, linkcolor=blue]{hyperref}
\usepackage{ae,aecompl}
\usepackage{amsmath}	
\usepackage{amssymb}	
\usepackage{dirtytalk}

\newcommand{\Msun}{ h^{-1}{\rm M_{ \odot}}}

\newcommand{\ihMpc}{ h\,{\rm Mpc}^{-1}}

\newcommand{\code}{\tt{}}
\newcommand{\mycomment}[1]{}

\begin{document}
\title{DES Y3 cosmic shear down to small scales: constraints on cosmology and baryons}

\author{Giovanni Aric\`o
          \inst{1,2}\fnmsep\thanks{E-mail:giovanni.arico@uzh.ch (GA)},
          Raul E. Angulo\inst{2,3}\fnmsep\thanks{E-mail:reangulo@dipc.org (REA)},
            Matteo Zennaro\inst{2,4},
            Sergio Contreras\inst{2}, Angela Chen\inst{5}, \\
 \and Carlos Hernández-Monteagudo\inst{6,7}}

   \institute{
Institute for Computational Science, University of Zurich, Winterthurerstrasse 190, 8057 Zurich, Switzerland \and
Donostia International Physics Center (DIPC), Paseo Manuel de Lardizabal, 4, 20018, Donostia-San Sebasti\'an, Guipuzkoa, Spain\and
IKERBASQUE, Basque Foundation for Science, 48013, Bilbao, Spain.\and
Department of Physics (Astrophysics), University of Oxford, Denys Wilkinson Building, Keble Road, Oxford, OX1 3RH, UK\and
Kavli Institute for the Physics and Mathematics of the Universe (WPI), UTIAS, The University of Tokyo, Kashiwa, Chiba 277-8583, Japan\and
Department of Astrophysics Research, Instituto de Astrof\'{i}sica de Canarias, C.\/V\'{i}a L\'{a}ctea, s\/n, E-38205, La Laguna, Tenerife, Spain\and
Departamento de Astrof\'{i}sica, Universidad de La Laguna, Avenida Francisco S\'{a}nchez, s\/n, E-38205, La Laguna, Tenerife, Spain
}

\date{Received XXX; accepted YYY}

\abstract{
We present the first analysis of cosmic shear measured in DES Y3 that employs the entire range of angular scales in the data. To achieve this, we build upon recent advances in the theoretical modelling of weak lensing provided by a
combination of $N$-body simulations, physical models of baryonic processes, and neural networks. Specifically, we use {\tt BACCOemu} to model the linear and nonlinear matter power spectrum including baryonic physics, allowing us to robustly exploit scales smaller than those used by the DES Collaboration. We show that the additional data produce cosmological parameters that are tighter but consistent with those obtained from larger scales, while also constraining the distribution of baryons.
In particular, we measure the mass scale at which haloes have lost half of their gas, $\log\,M_{\rm c}=14.38^{+0.60}_{-0.56}\, \log (\Msun)$, and a parameter that quantifies the weighted amplitudes of the present-day matter inhomogeneities, $S_8=0.799^{+0.023}_{-0.015}$. Our constraint on $S_8$ is statistically compatible with that inferred from the Planck satellite's data at the $0.9\sigma$ level. We find instead a $1.4\sigma$ shift in comparison to that from the official DES Y3 cosmic shear, because of different choices in the modelling of intrinsic alignment, non-linearities, baryons, and lensing shear ratios. 
We conclude that small scales in cosmic shear data contain valuable astrophysical and cosmological information and thus should be included in standard analyses.}


\titlerunning{DES Y3 cosmic shear down to small scales}
\authorrunning{G.Aricò et al.} 

\maketitle


\section{Introduction}
\label{sec:intro}
The standard cosmological model, $\Lambda {\rm CDM}$, has been heavily stress-tested in the last decade. The increasingly precise measurements of the Cosmic Microwave Background (CMB), and the recently achieved competitivity of large-scale-structure (LSS) missions, have highlighted some tensions on cosmological parameters estimated by different probes. In particular, a significant tension ($>4\sigma$) is found in the value of the expansion rate of the Universe today, as estimated using the inverse ladder or the CMB data from the Planck Satellite \citep[the so-called $H_0$ tension, see e.g.][]{Verde2019,Riess2022}. A milder tension (2-3$\sigma$) is found when constraining the growth of structure with CMB data and low-redshift probes such as weak gravitational lensing (WL) and galaxy clustering \citep[the so-called $\sigma_8$ or $S_8$ tension, see e.g. ][]{Heymans2021,DESY3_3x2pt,Garcia2021}.

Many ideas have been proposed to solve or at least relieve these tensions: exotic models of dark energy \citep[e.g.][]{Pourtsidou&Tram2016,Marra2021,Heisenberg2023} and dark matter \citep[e.g. decaying dark matter, see][]{Chen2021,Bucko2022}, and interacting dark sector \citep[e.g.][]{Lucca2021}, modified gravity \citep[see e.g.][]{Nguyen2023}, a non-linear suppression of the matter power spectrum, possibly given by baryonic physics \citep{Schneider2022,Amon&Efstathiou2022}, and modifications to the halo mass function \citep{Gu2023}. However, no definitive consensus has been reached up to now \citep{Verde2019,Wong2020}. Although these tensions could be caused by a fundamental shortcoming of $\Lambda$CDM, they could also originate from physical processes absent in the theory modelling or from unknown systematic errors in the observations. 

In the spirit of advancing the current state of cosmological analyses by adopting a more complete description of cosmic probes,
we present a reanalysis of the cosmic shear measured by the Dark Energy Survey (DES) \citep{DES2005,DES2016,Secco2022,Amon2022}. We include several improvements in the theoretical modelling of the data, most notably an explicit model for the role of baryons in WL, which allow us to include small scales usually neglected in previous analyses. With these, we will address the issue of to what degree the current $S_8$ tension could be caused by baryonic physics.

Cosmic shear is the correlation in the apparent shape of galaxies induced by the gravitational potential of the intervening matter between those galaxies and us. Shear is particularly interesting because it directly probes the cosmic density field (dark matter and baryons) bypassing the need for modelling galaxy bias. Thus, it offers a complementary probe to galaxy survey analyses.

The cosmic shear signal is very weak, therefore only large photometric surveys so far have had enough statistical power to competitively constrain cosmological parameters. The analysis is quite complex and relies on robust modelling of galaxy shapes, photometric redshifts, non-linearities in the growth of structures, intrinsic alignment of galaxies, etc. 
Nonetheless, in the last years several Collaborations, e.g. DES \cite{DESY3_3x2pt}, KiDS \citep{Heymans2021}, and HSC \citep[][]{hsc2019}, have successfully carried out these kinds of analyses. In the next years, {\it stage IV} surveys e.g. Euclid \citep{euclid}, DESI \citep[][]{DESI}, LSST \citep[][]{LSST2019}, are expected to dramatically improve on the current cosmological constraints.

To fully exploit current and future WL data, we need to model carefully all the physical processes that shape the distribution of matter in the universe. In particular, astrophysical feedback, e.g. supermassive black hole accretion and supernovae feedback, pushes gas outside dark matter haloes, modifying the cosmic matter density fields on small scales \citep{Schaye2010,Schneider&Teyssier2015,vandaalen2020}. 
This causes a non-trivial suppression of the expected cosmic shear signal, depending on the strength of the feedback and also from cosmology, mainly via the relative quantity of baryons available $\Omega_{\rm b}/\Omega_{\rm m}$ \citep{Schneider&Teyssier2019,vandaalen2020,Arico2020c}. Ignoring the role of baryonic processes is known to bias cosmological inferences and it has been identified as one of the main WL systematics, especially when considering the WL signal on small scales \citep[see e.g.][]{Semboloni2011,Chisari2018,Schneider2020}. 

Several approaches have been proposed to mitigate the effects of baryons on cosmic shear, including analytical parameterizations, \citep[e.g][]{HarnoisDeraps2015}, Principal Components Analysis (PCA) \citep{Eifler2015,Huang2019}, halo model \citep[e.g.][]{Semboloni2011,Mohammed2014,Fedeli2014,Mead2015,Debackere2019, Mead2020,Mead2020b}, and baryonification \citep{Schneider&Teyssier2015,Schneider&Teyssier2019,Arico2020}. 

The official DES analysis does not attempt to model baryons, but instead, it relies on angular scale cuts informed by hydrodynamical simulations, where data points at scales believed to be potentially affected by baryons are discarded. 
In this way, only 227 out of the total 400 data points are used for the analysis \footnote{\cite{Secco2022,Amon2022} also employ another angular scale cut, referred to as \say{$\Lambda {\rm CDM}$ optimised}, with 273 data points left. For simplicity, in this work we refer only to the \say{standard} DES scale cuts.}, thus not using all the available information. We take a different approach in this work, aiming to directly model the relevant baryonic processes. This has the advantage of fully exploiting the DES data, and also is expected to provide more conservative cosmological estimates since they i)  will be obtained after marginalisation over the uncertainty associated with baryons and ii) do not make any a priori assumption about the range of scales affected by baryons. 

In our analysis, we account for the effects of galaxy formation and gas physic using a baryonification algorithm \citep{Arico2020, Arico2020b} on top of the outputs of cosmological $N$-body simulations. Baryonification displaces the particles in gravity-only simulations according to analytical prescriptions based on physically-motivated assumptions. This approach is flexible enough to match at a per cent level the modifications induced by baryons in 2 and 3-point statistics as measured in tens of different state-of-the-art hydrodynamical simulations. Operationally, we use the neural-network emulators collected in {\tt BACCOemu} \citep{Angulo2020,Arico2020c,Arico2021}, which deliver accurate and fast predictions of the linear, non-linear, and baryonic contribution to the matter power spectrum. We simultaneously vary the parameters of the cosmological and baryonic model, thus capturing possible degeneracies. We find that in this way we are able to further extract cosmological information and separate it from the astrophysical content. 

This paper is structured as follows: in \S\ref{sec:data} we briefly summarise our dataset, the DES Y3 cosmic shear catalogue; in \S\ref{sec:model} we describe the details of our model and pipeline; in \S\ref{sec:analysis} we report our choices for the Bayesian inference; in \S\ref{sec:S8_tension} we compare our results against previous works and external datasets in light of the $S_8$ tension; in \S\ref{sec:conclusions} we discuss our results and give our conclusions. 

\begin{figure*}
\includegraphics[width=0.98\textwidth]{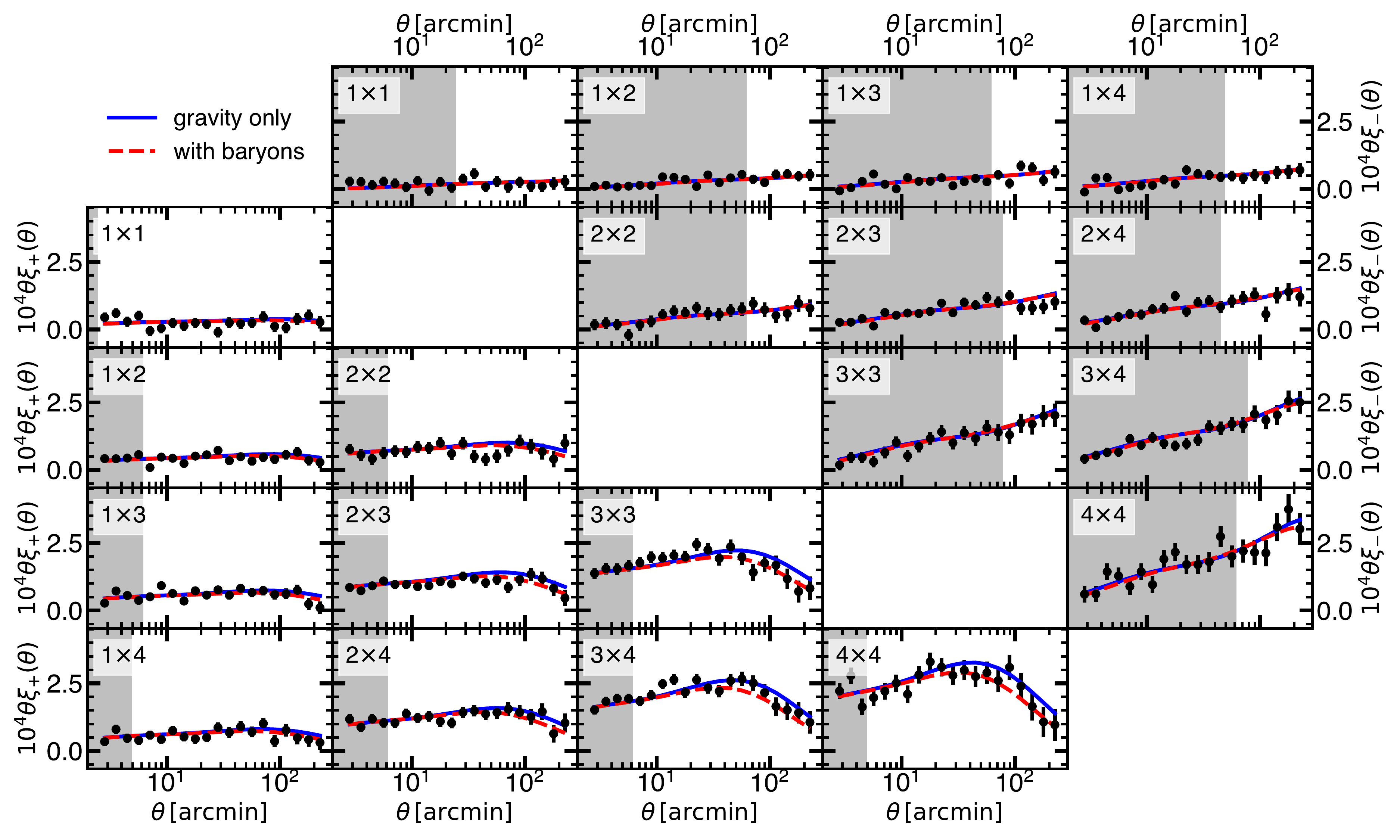}
\caption{ 
Cosmic shear correlation functions, $\xi_{\pm}$, measured by DES Y3 in 4 redshift bins, for a total of $400$ data points. In the lower-left corner we display $\xi_{+}$, whereas in the upper-right corner we show $\xi_{-}$, both multiplied by $10^{4} \theta$ for clarity. The corresponding redshift bins are indicated on each subplot. The grey bands show the scales discarded in the analysis carried out by the DES Collaboration. We display the best-fitting models we obtain using the full data vector when considering only gravity (blue solid line), and additionally considering baryonic processes (red dashed line).
}
\label{fig:datavector}
\end{figure*}

\section{Data}
\label{sec:data}

We employ the public comic shear dataset released by the DES Collaboration after 3 years of data collection\footnote{\url{https://desdr-server.ncsa.illinois.edu/despublic/y3a2_files/datavectors/}}. 
This includes the cosmic shear correlation functions measured using the \say{{\tt GOLD} catalogue}, which includes more than 100 million galaxy shapes, and the corresponding  covariance matrix. 
The sky coverage is $4143 \, {\rm deg}^2$, the mean redshift of the source galaxies is $z=0.63$ and the galaxy weighted number density is $n_{\rm eff} = 5.59 \, {\rm arcmin}^{-2}$. 
We refer to the DES papers for more details \citep{Flaugher2015,Morganson2018,Sevilla2021,Gatti2021,Myles2021,Krause2021,Amon2022,Secco2022}.

The catalogue of source galaxies is split into 4 broad redshift bins, roughly equi-populated between $z=[0,1.5] $ (the maximum nominal redshift is at $z=3$, although after $z=1.5$ the number density is very low). The 2 shear correlation functions, $\xi_{\pm}$, are measured in 20 logarithmic angular bins between 2.5 and 250 arcminutes. Joining auto and cross-correlations for both $\xi_+$ and $\xi_-$, we have a total of 400 data points, reduced to 227 when considering DES scale cuts. 
The covariance is obtained analytically by summing the Gaussian contribution, the 4-point connected part, the super-sample contribution, the survey geometrical effects, and the shape noise \citep{Friedrich2021}.
We show our data vector in Fig.~\ref{fig:datavector}, where we highlight with grey bands the data points removed by the DES scale cuts.

\section{Model}
\label{sec:model}

\subsection{Cosmic shear}
We obtain the shear correlation functions in the redshift bins $i,j$  $\xi^{i,j}_{\rm \pm}(\theta)$ by decomposing the shear field in E and B modes and combining the angular power spectra as 

\begin{equation}
\begin{split}
\xi^{i,j}_{\rm \pm}(\theta) = \sum\limits_{\ell} \frac{2\ell +1}{2 \pi \ell ^2 \, (\ell +1)^2} \left[ G^{+}_{\ell,2} (\cos \theta) \pm G^{-}_{\ell,2}(\cos \theta) \right] \\ 
\times
\left[ C^{i,j}_{\rm EE}(\ell) \pm C^{i,j}_{\rm BB}(\ell) \right]
\end{split}
\label{eq:xi}
\end{equation}
 \citep{Crittenden2002,Schneider2002,Krause2021}. The $G^{\pm}_{\ell,2}$ functions are defined from the Legendre polynomials following \cite{Stebbins1996}. We compute efficiently Eq.\ref{eq:xi} by using the Fast Fourier Transforms {\tt FFTLog} \citep{Talman1978}. 
 We use the Limber approximation \citep{Limber1953}, which is shown to be accurate enough for DES Y3 data \citep{Krause2021}. We have explicitly checked that the post-Limber correction proposed by \cite{Kitching2017} does not impact significantly our results ($\le 0.1\sigma$ in the cosmology inference).

The angular power spectra $C_{\rm EE}^{i,j}(\ell), C_{\rm BB}^{i,j}(\ell)$ are computed summing the relevant contributions, i.e. 

\begin{equation}
C_{\rm EE}^{i,j}(\ell) = C_{\rm GG}^{i,j}(\ell) + C_{\rm II, EE}^{i,j}(\ell) + C_{\rm GI}^{i,j} + C_{\rm IG}^{j,i},
\end{equation}

\begin{equation}
C_{\rm BB}^{i,j}(\ell) = C_{\rm II, BB}^{i,j}(\ell).
\end{equation}

The purely gravitational signal is given by the term $C_{\rm GG}(\ell)$, which only contributes to E modes. Within a flat $\Lambda$CDM model, we can compute this term as:

\begin{equation}
C^{i,j}_{\rm GG}(\ell) = \int_{0}^{\chi_{\rm H}}  \frac{g^i(\chi) g^j(\chi)}{\chi^2} P_{\rm GG} \left( \frac{\ell + 1/2}{\chi}, z(\chi) \right) d\chi,
\end{equation}

\noindent where $P_{\rm GG}(k,z)$ is the matter power spectrum, $\chi$ is the comoving distance and $z$ the corresponding redshift. The lensing kernel $g^i(\chi)$ reads

\begin{equation}
g^i(\chi) = \frac{3}{2}\Omega_{\rm m} \left( \frac{H_0}{c} \right)^2  \frac{\chi}{a} \int_{z(\chi)}^{z_{\rm H}} n^i(z') \frac{\chi(z\prime)-\chi(z)}{\chi(z\prime)}dz\prime,
\end{equation}

\noindent where $n^i(z)$ is the normalised galaxy redshift distribution in each $i$ bin \citep[see ][for details on how these are obtained in the DES case]{Myles2021}, and $z_{\rm H}$ is the redshift of the Hubble sphere.

The other terms, $C_{\rm II, EE}(\ell)$ and $C_{\rm GI}$, are given by the intrinsic alignment of galaxies, and we recap them in the next section.

\subsection{Intrinsic alignment}
\label{sec:ia_define}
The measured shear signal includes the physical correlation of galaxy shapes in the sky, often referred to as the intrinsic alignment of galaxies. Within a flat $\Lambda$CDM framework and using the Limber approximation, the auto-correlation of the intrinsic alignment is given by

\begin{equation}
C^{i,j}_{II}(\ell) = \int_{0}^{\chi_{\rm H}}  \frac{n^i n^j}{\chi^2} P_{\rm II} \left( \frac{\ell+ 1/2}{\chi}, z \right) d\chi,
\end{equation}

\noindent and the cross-correlation between gravitational shear and intrinsic alignment reads

\begin{equation}
C^{i,j}_{GI}(\ell) = \int_{0}^{\chi_{\rm H}} 
\frac{g^i n^j + g^j n^i }{\chi^2} P_{\rm GI} \left( \frac{\ell+ 1/2}{\chi}, z \right) d\chi, 
\end{equation}

\noindent where for brevity we have omitted the dependence of $g,n$, and $z$ on $\chi$.

We implement two models of the intrinsic alignment of galaxies: the non-linear model \citep[NLA,][]{Bridle&King2007}, and the more complex tidal alignment \& tidal torque model \citep[TATT,][]{Blazek2019}. The NLA model, with two free parameters, can be seen as a specific case of the more general TATT (5 free parameters). 

Within NLA, the intrinsic alignment auto-correlation, and cross-correlation with cosmic shear are, respectively,

\begin{equation}
P_{\rm II} (k, z) = A_1(z)^2 P_{\rm GG}(k,z),
\end{equation}
\begin{equation}
P_{\rm GI} (k, z) = A_1(z) P_{\rm GG}(k,z), 
\end{equation}

with 

\begin{equation}
A_1 (z) = - a_1 C_1 \frac{\rho_{\rm crit} \Omega_{\rm m}}{D(z)} \left( \frac{1+z}{1+z_0} \right)^{\eta_1}.
\end{equation}

Here, $C_1$ is a normalization constant typically set to $C_1 = 5 \times 10^{-14} \, (h^2 {\rm M_{\odot} Mpc^3})^{-2}$ \citep{Hirata&Seljak2004,Bridle&King2007}, $z_0=0.62$ is typically assumed in DES analysis \citep{Secco2022,Amon2022}, and $a_1$, $\eta_1$ are free parameters. 

In the TATT model, we have instead  

\begin{equation}
\begin{split}
P_{\rm II,EE} = A_1^2 P_{\rm GG} + 2 A_1 A_{1,\delta} P_{\rm 0|0E} + A_{1,\delta}^2 P_{\rm 0E|0E} \\
 + A_2^2 P_{\rm E2|E2}+2 A_1 A_2 P_{0|E2} 
+ 2 A_{1,\delta} A_2 P_{0E|E2}, 
\end{split}
\end{equation}

\begin{equation}
P_{\rm II,BB}  = A_{1,\delta}^2 P_{\rm 0B|0B} + A_2^2 P_{\rm B2|B2} + 2 A_{1,\delta} A_2 P_{0B|B2}, 
\end{equation}

and

\begin{equation}
P_{\rm GI} = A_1 P_{\rm GG} + A_{1,\delta} P_{\rm 0|0E} + A_2 P_{0|E2},
\end{equation}

\noindent where for the sake of brevity we have omitted scale and redshift dependencies. The power spectra $P_{\rm 0|0E}, P_{\rm 0E|0E},P_{\rm 0B|0B}$, etc.,  are computed within perturbation theory in \cite{Blazek2019}, to which we refer for the details. We evaluate these power spectra using the public code {\tt FAST-PT} \citep{McEwen2016,Fang2017}.

Following the convention of \cite{Secco2022,Amon2022}, we define 
\begin{equation}
A_2 (z) = 5 a_2 C_1 \frac{\rho_{\rm crit} \Omega_{\rm m}}{D^2(z)} \left( \frac{1+z}{1+z_0} \right)^{\eta_2},
\end{equation}
\begin{equation}
A_{1,\delta} (z) = b_{\rm TA} A_1 (z).
\end{equation}

We therefore have the additional free parameters $a_2$, $\eta_2$, $b_{\rm TA}$. When fixing $a_2=0$ and $b_{\rm TA}=0$, TATT reduces to NLA.

There is no consensus on the regime of validity of NLA and TATT \citep[see e.g.][]{Samuroff2022}. Previous works on DES data have constrained the amplitude parameters of the intrinsic alignment, $a_1$,$a_2$, of the \say{GOLD catalogue} to be consistent with zero \citep{Secco2022,Amon2022,DESY3_3x2pt,Samuroff2022}. Indeed, Bayesian evidence prefers a model with no intrinsic alignment at all, followed by the NLA model. TATT is disfavoured because of its relatively high number of parameters combined with a low signal compared to the cosmological one. This can be seen as a preference towards simpler intrinsic alignment models for DES data. Therefore, in this work, we employ NLA as our fiducial model. Nonetheless, since we are employing angular scales previously discarded, we repeat the full analysis with TATT to check the robustness of our inference.

\subsection{Matter power spectrum}

We evaluate the matter power spectrum $P_{\rm GG}(k,z)$ employing a series of Neural Network emulators from the BACCO Simulation project \citep{Angulo2020}. Specifically, the matter power spectrum is decomposed into three different components: a linear part given by perturbation theory, a non-linear boost function given by $N$-body simulations, and a baryonic correction given by a baryonification algorithm. 

The linear component is a direct emulation of the Boltzmann solver CLASS \citep{Lesgourgues2011}, which speeds up the calculations by several orders of magnitude \citep{Arico2021} while introducing a negligible error. The non-linear boost function is built by interpolating results at more than 800 different cosmologies, obtained from 5 high-resolution $N$-body simulations of $\approx 2 {\rm Gpc}$ and $4320^3$ particles, together with the methodology developed by \cite{A&W2010,Angulo&Hilbert2015,Zennaro2019,Angulo2020,Contreras2020}. This algorithm manipulates the output of a simulation to mimic the expected particle distribution in a very wide cosmological space, with an accuracy of $3\%$ in the power spectrum  at $k \sim 5 \ihMpc$ in $\Lambda$CDM including massive neutrinos \citep{Contreras2020,Angulo2020}. Finally, the baryonic correction is computed by applying a baryonification algorithm \citep{Schneider&Teyssier2015, Arico2020} to the $N$-body simulations. 

These emulators have been collected into the public repository {\tt BACCOemu} \citep{Angulo2020, Arico2020c, Arico2021}. 
Here, we use an updated version of the public emulators, which extends the non-linear boost functions from scales $k\le 5\ihMpc$ to $k \le 10\ihMpc$ and from redshifts $z\le 1.5$ to $z\le 3$ by employing a suite of 5 higher-resolution $N$-body simulations ($L\approx 750 {\rm Mpc}$ and $2288^3$ particles). 
The new emulator also features an updated version of the cosmology rescaling algorithm, including a new halo mass function and concentration-mass relation \citep{Ondaro2021,LopezCano2022} which improves its accuracy. Moreover, the cosmological parameter space has been expanded thanks to the addition of a suite of 35 new simulations, so that encapsulate the priors used here. The only exception is the cold matter cosmic density, $\Omega_{\rm c}$, which nonetheless has been extended from $[0.23,0.4]$ to $[0.15,0.47]$ (we extrapolate with {\code HALOFIT} outside of these boundaries \footnote{Typically, more than $99\%$ of the posterior evaluations are within ${\code BACCOemu}$ priors.}). We show a comparison between {\code HALOFIT} and {\code BACCOemu} in App.~\ref{app:nl_info}, and we refer to Zennaro et al. (in prep.) for further details and validation. 

\begin{table}
\begin{tabular}[t]{lccc}
\hline
\multicolumn{2}{c}{\textbf{Cosmology}}\\
$\Omega_{\rm m}$& $[0.1,0.7]$\\
$A_{\rm s}$ & $[5\times 10^{-10}, 5\times 10^{-9}]$\\
$h$& $[0.55,0.90]$\\
$\Omega_{\rm b}$& $[0.03,0.07]$\\
$n_{\rm s}$& $[0.87,1.07]$\\
$M_{\nu}$& $[0.0559,0.400]$ eV\\
\hline
\multicolumn{2}{c}{\textbf{Baryons}}\\
$\log{M_{\rm c}}$ & $[9.0,15.0]$ $\log(\Msun)$\\
$\log{\eta}$ & $[-0.7,0.7]$\\
$\log{\beta}$ & $[-1.0,0.7]$\\
$\log M_{z0,\rm cen}$ & $[9.0,13.0]$ $\log(\Msun)$\\
$\log{\theta_{\rm inn}}$ & $[-2.00, -0.53]$\\
$\log{\theta_{\rm out}}$ & $[-0.48, 0.00]$\\
$\log{M_{\rm inn}}$ & $[9.0, 13.5]$ $\log(\Msun)$\\
\hline
\multicolumn{2}{c}{\textbf{Intrinsic Alignment}}\\
$a_1$ & $[-5, 5]$\\
$\eta_1$ & $[-5, 5]$\\
$a_2$ & $[-5, 5]$\\
$\eta_2$ & $[-5, 5]$\\
$b_{\rm ta}$ & $[0, 2]$\\
\hline
\multicolumn{2}{c}{\textbf{Photo-z shift}}\\
$\Delta z_{\rm s}^1$ & $\mathcal{N}(0.000, 0.018)$\\
$\Delta z_{\rm s}^2$ & $\mathcal{N}(0.000, 0.015)$\\
$\Delta z_{\rm s}^3$ & $\mathcal{N}(0.000, 0.011)$\\
$\Delta z_{\rm s}^4$ & $\mathcal{N}(0.000, 0.017)$\\
\hline
\multicolumn{2}{c}{\textbf{Shear calibration}}\\
$m^1$ & $\mathcal{N}(-0.0063, 0.0091)$\\
$m^2$ & $\mathcal{N}(-0.0198, 0.0078)$\\
$m^3$ & $\mathcal{N}(-0.0241, 0.0076)$\\
$m^4$ & $\mathcal{N}(-0.0369, 0.0076)$\\
\hline
\end{tabular}
\centering
\caption{Priors on the free parameters employed in our Bayesian analyses.  
The intrinsic alignment parameters $a_2$, $\eta_2$, $b_{\rm ta}$ are fixed to 0 in our fiducial run. We also impose a prior on the combination of parameters $\Omega_{\rm b} h^2 \in [0.009,0.040]$.}   
\label{tab:priors}
\end{table}%

\subsection{Baryonic effects}
\label{subsec:bcm}
We model the baryonic processes that impact the cosmic density field with a baryonification scheme \citep{Schneider&Teyssier2015,Arico2020}. 
The baryonification, or Baryon Correction Model (BCM), displaces the particles of a gravity-only simulation according to analytical corrections to take into account the effects of different baryonic processes on the density field. In this framework, haloes are assumed to be constituted by galaxies, gas in hydrostatic equilibrium, and dark matter. 
A given fraction of the gas is expelled from the haloes by accreting supermassive black holes, and the dark matter back-reacts on the baryon gravitational potential with a quasi-adiabatic relaxation. 
The BCM has been proven flexible enough to reproduce the 2-point and 3-point statistics of several hydrodynamical simulations \citep{Arico2020b,Giri&Schneider2021}, and has been used to analyse cosmic shear data \citep{Schneider2022,Chen2023}. 

In this work, we employ the emulator of the BCM suppression in the matter power spectrum described in \cite{Arico2020}. The emulator fully captures the degeneracies between astrophysical processes and cosmology, while being accurate at per cent level \citep{Arico2020c}. It has a total of 15 free parameters, 7 to describe the baryonic processes, plus 8 for cosmology. We further test the accuracy of the BCM emulator in the  modelling of DES Y3 cosmic shear analysis in App.~\ref{app:hydro_compa}, where we compare the BCM against the hydrodynamical simulation BAHAMAS \citep{McCarthy2017,McCarthy2018}. 

\begin{figure}
\includegraphics[width=0.4\textwidth]{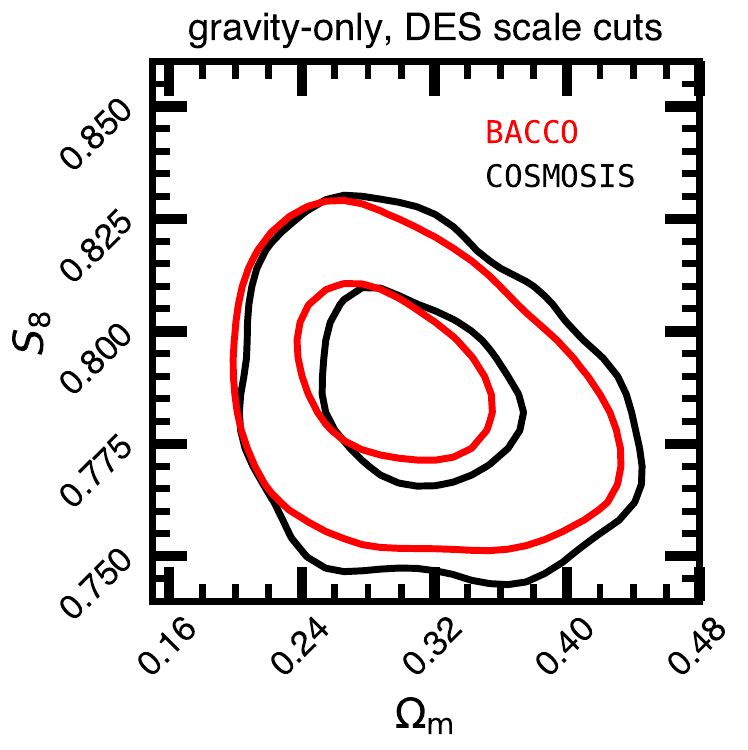}
\caption{Comparison between the posterior distribution on $S_8\equiv \sigma_8\sqrt{\Omega_{m}/0.3}$ and $\Omega_m$ obtained using {\code CosmoSIS} (employed by the DES Collaboration) and our pipeline, embedded in the {\code BACCO} library. For this comparison, both cases compute the nonlinear matter power spectrum using {\code HALOFIT}. The intrinsic alignment model used is NLA, without the contribution of shear ratios. Here and throughout this paper, we always show $1\sigma$ and $2\sigma$ credible levels (we note that DES Collaboration papers typically display $68\%$ and $95\%$ credible levels instead).}
\label{fig:halofit_DES}
\end{figure}

The cosmological parameter space in which the BCM emulator has been trained is smaller with respect to the DES priors. Thus, when sampling a cosmological model that lies outside the emulator space, we set the cosmological parameters of the baryonic response to the closest cosmology available. 
Additionally, we explicitly set the (cold) cosmic baryon fraction, $\Omega_{\rm b}/\Omega_{\rm cdm + b}$, to the closest available value. We expect this to be a good approximation because  i) the $1-\sigma$ region obtained analysing only the large scales of DES broadly fits into the emulator priors; ii) the baryonic feedback is mainly dependent on the (cold) cosmic baryon fraction, which is reasonably well covered by our emulator ($\Omega_{\rm b}/\Omega_{\rm cdm + b} \in [0.1,0.26]$, more than $98\%$ of the posterior samples in our fiducial run falls within this interval).
 
Our BCM emulator has been previously applied to DES Y3 data in \cite{Chen2023} to constrain the impact of astrophysical feedback. They used small-scales DES Y3 shear measurements to constrain $M_{\rm c}$ -- the baryonic parameter data is most sensitive to. That is, the characteristic halo mass ($M_{200, \rm crit}$ expressed in $\Msun$) in which half of the cosmic gas fraction is expelled from the halo by astrophysical processes. \cite{Chen2023} find that $\log M_{\rm c}=14.12^{+0.62}_{-0.37}$. In the analysis, the cosmology was varied within a prior given by the posterior provided by the 3x2pt analysis of DES Y3, that is the combination of cosmic shear with galaxy clustering and galaxy-galaxy lensing \citep{DESY3_3x2pt}. 

The other free parameters of the BCM describe the shape of the density profile of the hot gas ($\theta_{\rm inn}$,$\theta_{\rm out}$,$M_{\rm inn}$), the galaxy-halo mass ratio ($M_{z0,\rm cen}$), the AGN feedback range $\eta$, and the gas fraction - halo mass slope ($\beta$). We refer the reader to \cite{Arico2020b,Arico2020c} for further details. 

Here, we set free all the BCM parameters to avoid relying on a specific hydrodynamical simulation. 
We also explicitly test the impact of fixing all the BCM parameters but $M_{\rm c}$, as done in \cite{Chen2023}.

\begin{table*}
    \begin{tabular}{c c c c c}
    Model & \multicolumn{2}{c}{$\chi^2/{d.o.f.(N_{\rm free})}$} & \multicolumn{2}{c}{$\chi^2/{d.o.f.(N_{\rm eff})}$} \\ \hline

     &  DES scale cuts & all scales & DES scale cuts & all scales \\ \hline
    NLA BCM7 (fiducial) & 226.98/204=1.11 & 414.14/377=1.1 & 226.98/224.69=1.01 & 414.14/397.16=1.04\\
    NLA GrO & 227.98/211=1.08 & 416.46/384=1.08 & 227.98/224.63=1.01 & 416.46/397.37=1.05 \\
    NLA BCM1 & 227.58/210=1.08 & 414.23/383=1.08 & 227.58/224.61=1.01 & 414.23/397.11=1.04\\
    NLA BCM-extreme & 227.24/211=1.08 & 419.03/384=1.09 & 227.24/224.6=1.01 & 419.03/397.77=1.05\\
    \hline
    TATT BCM7 & 226.28/201=1.13 & 411.22/374=1.1 & 226.28/224.14=1.01 & 411.22/396.21=1.04\\
    TATT GrO & 226.37/208=1.09 &  410.66/381=1.08 & 226.37/224.28=1.01 & 410.66/396.16=1.04\\
    TATT BCM1 & 225.28/207=1.09 & 407.65/380=1.07 & 225.28/223.87=1.01 & 407.65/396.35=1.03\\
    TATT BCM-extreme & 227.2/208=1.09 & 408.69/381=1.07 & 227.2/223.77=1.02 & 408.69/396.37=1.03\\
    \hline
    \end{tabular}
     \centering
    \caption{Goodness of fits of our models given the DES Y3 data vector. We calculate the degrees of freedom ($d.o.f.$) by subtracting the number of parameters $N_{\theta}$ from the number of data points ($N_{\mathcal{D}}$). We use as $N_{\theta}$ either the number of free parameters $N_{\rm free}$, or the number of effective parameters $N_{\rm eff}$ as defined in \protect\cite{Raveri&Hu2019}. We explore different models, with NLA or TATT intrinsic alignment, gravity-only, or applying a baryonification with 1 or 7 free parameters (BCM1 \& BCM7, respectively). We also include a model with an extreme implementation of AGN feedback.}
\label{tab:goodfit}
\end{table*}

\subsection{Nuisance Parameters}
\label{subsec:nuisance}
Following the DES Collaboration, we model the uncertainties on the photometric redshift estimate of source galaxies as a shift in the redshift distributions in each redshift bin $\Delta z_{\rm s}^i$. Furthermore, we model the unaccounted effects of the shape calibration and blending of the galaxy shapes with a multiplicative bias independent for each redshift bin, $m^i$.  For both, photometric errors and shear bias, we employ the informative priors used in \cite{Secco2022,Amon2022}.

\subsection{Pipeline}
For this work, we have implemented from scratch a WL analysis pipeline. The pipeline is interfaced with the cosmology library {\code BACCO} (Angulo et al., in prep.) and the {\code BACCOemu} emulators. It is written in {\code python}, with a multi-threaded {\code C} core for the most computationally-demanding functions.  As input, the pipeline takes a series of parameters (including cosmological, astrophysical, intrinsic alignment, photometric errors, and shear bias parameters), and it outputs the predicted shear correlation functions. 

We perform the Bayesian analysis with a nested sampling algorithm, the public code {\code POLYCHORD} \citep{Handley2015}. We follow the guidelines of \cite{Lemos2022} (Tab. 3) and use their "Publication quality" setting, which features 500 live points and a tolerance of 0.01. We assume a Gaussian likelihood with a covariance matrix provided by the DES collaboration. 

In contrast to the official DES analysis, we choose not to include the so-called \say{shear ratios} \citep{Sanchez2022,Amon2022}. Defined in \cite{Sanchez2022}, shear ratios measure the galaxy-galaxy lensing produced by the same lenses with different source galaxies samples. They found that, even at the smallest angular scales measured by DES Y3, these ratios are robust to changes in cosmology, baryonic processes, and galaxy bias, but highly sensitive to the source galaxy redshift distribution and intrinsic alignment. Despite the potential benefits, we opt for focusing only on shear data for two reasons: i) we avoid complications such as the modelling of galaxy bias, that we should coherently include in our current framework ii) we can better isolate the information on intrinsic alignment and redshift distributions coming from the small scales of cosmic shear. 

\begin{figure*}
\centering
\includegraphics[width=0.4\textwidth]{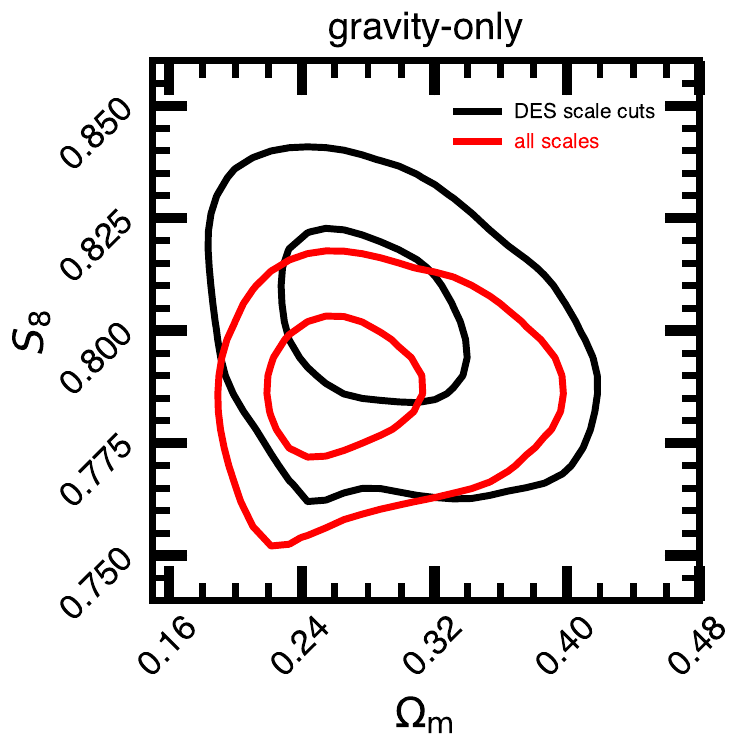}
\includegraphics[width=0.4\textwidth]{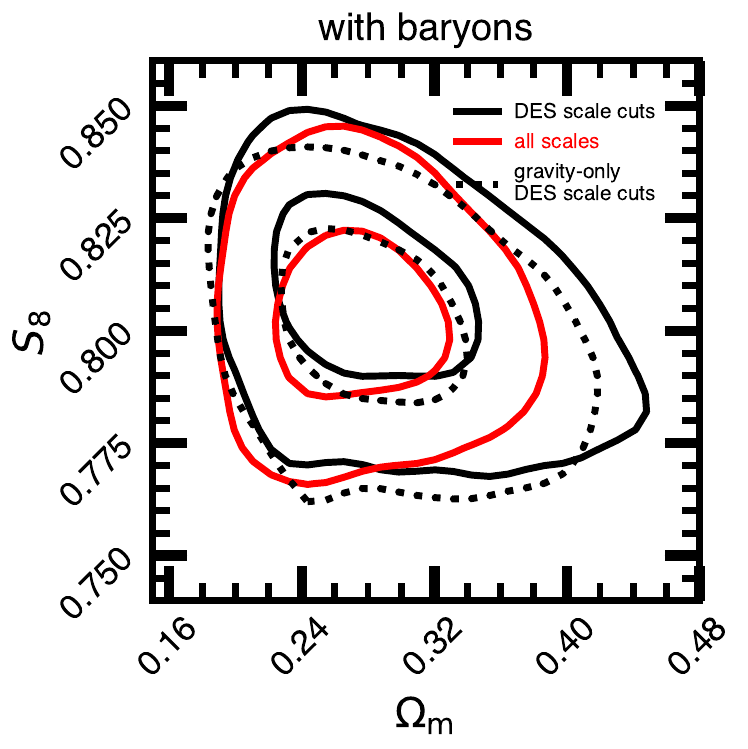}

\caption{$1\sigma$ and $2\sigma$ credible levels on $S_8$ and $\Omega_{\rm m}$ obtained using {\code BACCO}, considering only gravity in structure formation (left panel) and including baryonic effects (right panel). Black solid lines refer to runs with the application of scale cuts, and red solid lines to runs where all scales are used. The black dotted line in the right panel shows for reference the results of the gravity-only model with scale cuts.}
\label{fig:scale_cuts}
\end{figure*}

We generally use the same priors as the official DES analyses \citep{Secco2022,Amon2022,DESY3_3x2pt}, except for the upper boundary of $\Omega_{\rm m}$ which is 0.7 in our analysis instead of 0.9, of $h$ which is 0.9 instead of 0.91, and of the sum of neutrino masses $M_{\nu}$ that is 0.4 instead of 0.6, to fit in the parameter space of our emulator.  We have checked with {\tt HALOFIT} that these new priors do not significantly impact our final results. 
Specifically, we only detect a small bias of less than $0.1\sigma$ on the cosmological constraints, given by the different prior on $M_{\nu}$. This is in broad agreement with what was found in \citep{Secco2022,Amon2022} when fixing the sum on neutrino masses. The priors on baryonic parameters are those discussed in \cite{Arico2020c}, chosen to be wide enough to encompass X-ray observations and hydrodynamical simulations.

We summarise the priors we use in Tab.~\ref{tab:priors}. In our fiducial run, we use the NLA intrinsic alignment model, i.e. $a_2=\eta_2=b_{\rm ta}=0$. We also include an extra flat prior $\Omega_{\rm b} h^2 \le 0.04$ given by Big Bang Nucleosynthesis \citep{Beringer2012}, analogously to that adopted in {\code CosmoSIS} and by the DES Collaboration \citep{Zuntz2015,DESY3_3x2pt}.

We have validated our pipeline by performing a model comparison with the public codes implemented in {\code CosmoSIS} \citep{Zuntz2015} and Core Cosmology Library \citep[{\code CCL},][]{CCL2019}, finding an excellent agreement between the three. We have also performed a test run employing {\code HALOFIT} \citep{Takahashi2012} and NLA, and using the official DES scale cuts. We compare these results with the chains run with the DES pipeline and {\code CosmoSIS}\footnote{The chains we compare against are from private communication, since the public ones include shear ratios.} in Fig.~\ref{fig:halofit_DES}, where we show for convenience only the $\Omega_{\rm m}-S_8$ plane. Throughout the paper, we use the definition $S_8\equiv \sigma_8 \sqrt{\Omega_{\rm m}/0.3}$. We find a difference between the two pipelines of about $0.1\sigma$, a very good agreement when considering all the different details of the implementation (e.g. binning, numerical accuracy, interpolation, emulators, etc.). Although not shown here, we have checked that the agreement holds true for all the model's free parameters.  

As further validation of our pipeline, and in order to check for possible biases introduced when analysing the smallest scales used in this work, we compute the $\chi^2$ values obtained using an independent, state-of-the-art model, and compare them with ours. We use a model composed by i) {\code EuclidEmulator2} \citep{EuclidEmulator2} for the non-linear matter power spectrum; ii) the baryonic suppression measured in the BAHAMAS suite of hydrodynamical simulations \citep{McCarthy2017}; iii) the TATT intrinsic alignment of galaxies \citep{Blazek2019}.

Comparing this model with our fiducial one for different cosmological and baryonic parameters (Planck and DES Y3 best-fitting cosmologies \citep{Planck2018,Amon2022,Secco2022}, BAHAMAS low, reference, and high AGN suppressions), we find fractional differences in $\chi^2$ below 1\%. By contrast, the difference in $\chi^2$ between baryonic modelling and gravity-only is around 5-15\%. The difference in $\chi^2$ between NLA and TATT is highly dependent on the value of the TATT parameters ($a_2$, $b_{\rm TA}$), and can be lower than $1\%$ and larger than $20\%$. Apriori, we do not know the amplitudes of TATT for a survey like DES, although previous studies point toward low amplitudes (and thus smaller $\chi^2$ differences). Therefore, we compare a posteriori the results obtained with NLA and TATT in App.~\ref{app:ia_info}.

Finally, we note that with our pipeline we evaluate a likelihood in less than 0.5 seconds on a common laptop, whereas it takes about 8 seconds to run a standard DES Y3 cosmic shear likelihood evaluation with {\code CosmoSIS}. 

\section{Analysis}
\label{sec:analysis}

In this section, we employ the pipeline described and validated in \S\ref{sec:model} to obtain  constraints on cosmology, baryonic physics, and intrinsic alignment of galaxies. We explore different levels of complexity in our model and study their impact on cosmology inference. 
We will focus on the derived cosmological parameter $S_8$, and on $\Omega_{\rm m}$. 
When showing the credible levels in the $S_8-\Omega_{\rm m}$ plane, we display the $1\sigma$ and $2\sigma$ levels (i.e. 39\% and 86\% in 2D), as opposed to $68\%$ and $95\%$ normally shown in DES papers, to visually help the assessment of tensions between different models and datasets. 
Unless stated otherwise, we quote all our constraints as the mode of the 1D marginalised posterior, plus and minus the respective 34th percentiles. We caution against directly comparing with the official DES constraints, which typically report the mean of the marginalised posterior. To help the comparison, we have reanalysed the DES Collaboration chains with the same routine we use for our chains \citep[{\code ChainConsumer},][]{Hinton2016}. We report in Tab.~\ref{tab:model_params} the constraints obtained with the mode and in Tab.~\ref{tab:cum_params} with the median of the posteriors, and their respective 34th percentiles.      

\subsection{Constraints on cosmology and goodness of fit}
\label{subsec:cosmo_cons}

 In Fig.~\ref{fig:scale_cuts} we show the posterior distribution functions of the cosmological parameters $S_8$ and $\Omega_{\rm m}$ assuming only gravitational interactions (left panel) and including our fiducial 7-parameters baryonic model (right panel). In each case, we display our results when applying DES scale cuts (black lines) and when employing all scales (red). When using our fiducial model, and analysing all the angular scales available, we obtain $S_8=0.799^{+0.023}_{-0.015}$ and $\Omega_{\rm m}=0.252^{+0.066}_{-0.030}$. 
 In Tab.~\ref{tab:model_params} we report the constraints on $S_8$ and $\Omega_{\rm m}$ obtained with different modelling choices, whereas in App.~\ref{app:cosmo_posteriors} we show the posteriors of all the free cosmological parameters in our fiducial run.
 
 Both the models with and without baryons provide a statistically good fit of the data, as shown in Fig.~\ref{fig:datavector} (red solid and blue dash lines, respectively). More quantitatively, we report the goodness of the fit for all our models in Tab.~\ref{tab:goodfit}, defined with the reduced $\chi^2$:
 \begin{equation}
\hat{\chi}^2=\frac{\chi^2}{d.o.f.} = \frac{(\mathcal{D}-\mathcal{M})^{\rm T} \mathcal{C}^{-1} (\mathcal{D}-\mathcal{M})}{N_{\mathcal{D}}-N_{\theta}},
 \end{equation}
 
\noindent where $\mathcal{D}$ is our data vector, $\mathcal{M}$ is our model, $\mathcal{C}$ the covariance, and the degrees of freedom ($d.o.f.$) are found by subtracting the number of parameters $N_{\theta}$ from the number of data points ($N_{\mathcal{D}}$). We use as $N_{\theta}$ either the number of free parameters $N_{\rm free}$, or the number of effective parameters $N_{\rm eff}$ as defined in \cite{Raveri&Hu2019}. When using $N_{\rm free}$, we find a slightly higher $\hat{\chi}^2$ in the case with baryons with respect to the gravity-only ($\hat{\chi}^2=1.1$ and $\hat{\chi}^2=1.08$, respectively), due to the 7 extra free parameters. When using $N_{\rm eff}$ we find instead $\hat{\chi}^2=1.04$ with baryons and $\hat{\chi}^2=1.05$ in the gravity-only case, which corresponds to $p$-values of $0.27$ and $0.24$, respectively. Nonetheless, the posteriors in the $S_8$-$\Omega_{\rm m}$ plane are quite different, the gravity-only one being tighter and shifted towards lower values of $S_8$.        

\subsection{Cosmological information at small scales}
\label{subsec:cosmo_info}

Although the DES measurements of the shear correlation functions get to angles as small as $2.5$ arcminutes, the DES Collaboration has so far refrained from modelling such small scales because of possible biases in the cosmology inference due to the effects of baryons. In particular, they have set different angular scale cuts for $\xi_+$ and $\xi_-$ and each redshift bin. 

The DES scale cuts are chosen such that the difference in $\chi^2$ between analyses carried on with a given synthetic data vector and the same data vector contaminated with the baryonic effects predicted by the hydrodynamical simulation OWLS-AGN \citep{Schaye2010} is lower than a given threshold \citep[for more details, see e.g.][]{Krause2021}. This results in retaining 227 data points over 400 (166 in $\xi_+$, 61 in $\xi_-)$. 

To quantify the amount of cosmological information loss when discarding in the analysis the small angular separations, we run our pipeline with and without these scale cuts. First, we neglect the effects of baryonic physics to get a sense of the improvement in the cosmological constraints in an ideal scenario, even if likely the constraints will be biased. We show the $1\sigma$ and $2\sigma$ credible levels of $S_8$ and $\Omega_{\rm m}$ in the left panel of Fig.~\ref{fig:scale_cuts}.
Adopting the DES scale cuts, we obtain $S_8=0.802^{+0.019}_{-0.021}$, whereas including smaller scales we find $0.785^{+0.017}_{-0.012}$, a constraint $30\%$ tighter. This can be seen as the upper limit on the cosmological information that we can potentially gain when modelling the small scales. Note, however, that this figure is specific to DES Y3 data, since it depends on the statistical accuracy with which small scales are measured. For surveys with a higher number density of background galaxies, better photometry and angular resolution, we expect the gains to be more substantial.

When including small scales, we see how the $S_8$ posterior shifts by about $0.5\sigma$ towards lower values. This is likely caused by baryonic processes: the lack of modelling of the suppression in the matter power spectrum caused by baryons could be compensated by a lower inferred value of $\sigma_8$ and $\Omega_{\rm m}$.  Hence, to be able to exploit the data on small scales, it is necessary to explicitly model the role of baryonic physics.  

\begin{figure}
\includegraphics[width=0.4\textwidth]{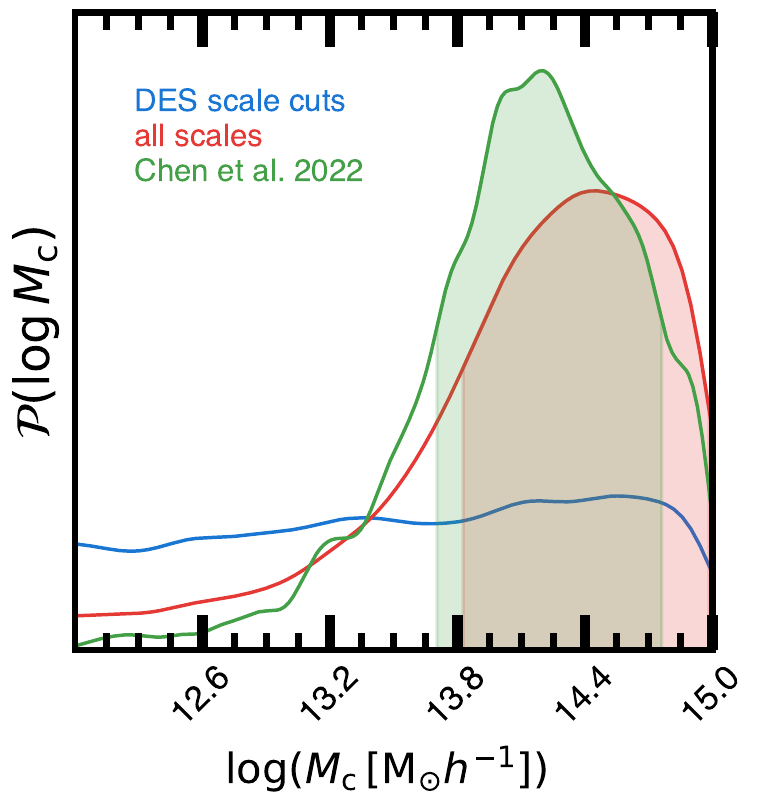}
\caption{Posterior distribution functions of $\log M_{\rm c}$ obtained in our fiducial run using all angular scales (red) and applying scale cuts (blue). We compare against the posterior obtained by \protect\cite{Chen2023} using only small scales, fixing 6 baryonic parameters and applying a prior on cosmology given by the 3x2pt analysis of DES Y3 \protect\citep{DESY3_3x2pt}.}
\label{fig:Mc}
\end{figure}

\begin{figure}
\includegraphics[width=0.44\textwidth]{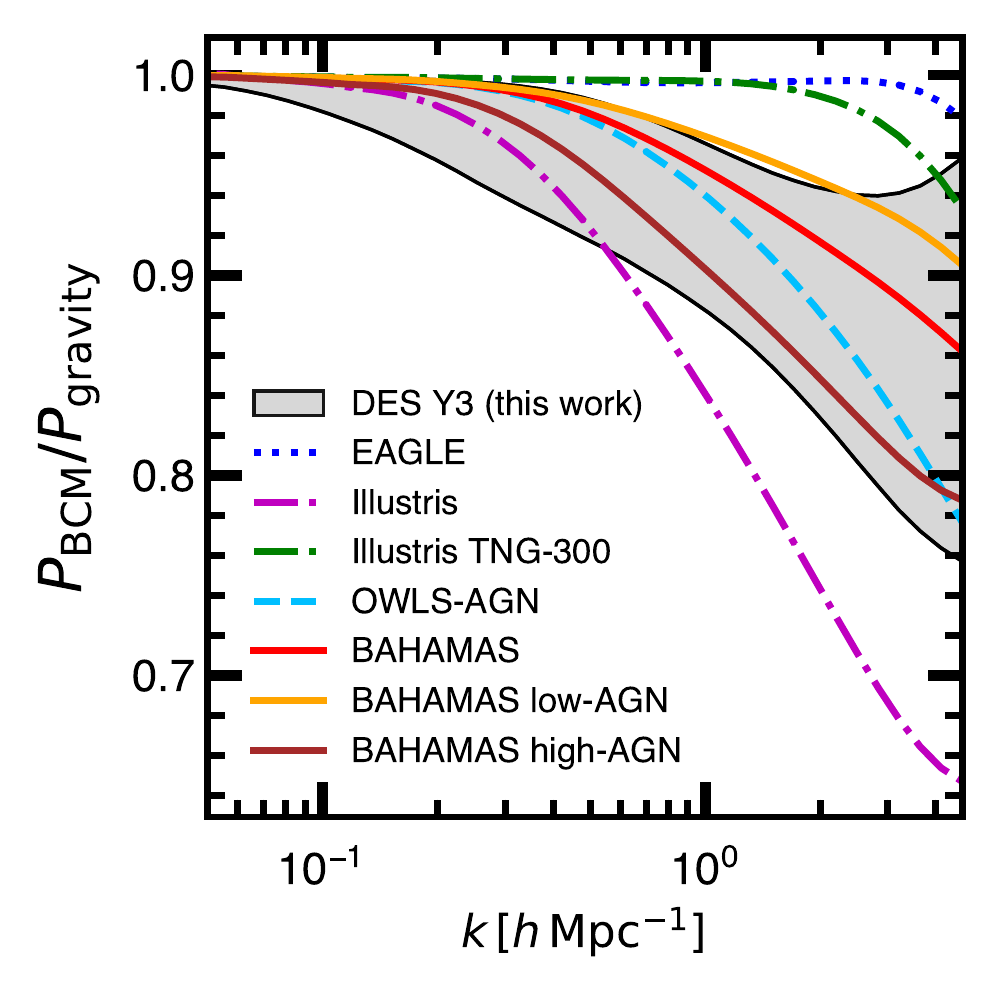}
\caption{$1\sigma$ credibility region of the suppression in the matter power spectrum at $z=0$ caused by baryons, inferred by the BCM model with DES Y3 data (grey shaded area). We display the ratio between the matter power spectrum affected by baryonic processes over the gravity-only one. For comparison, we show the BCM models obtained in \protect\cite{Arico2020c}, fitting several state-of-the-art hydrodynamical simulations referenced in the main text, according to the legend.}
\label{fig:Sk}
\end{figure}

\subsection{Impact of baryons}
\label{subsec:baryons_info}

Baryonic processes like gas cooling, galaxy formation, and active galactic nuclei (AGN) modify the cosmic matter power spectrum in a nontrivial way. The scales and amplitudes of their effects are currently debated, and can potentially affect the cosmology inference, if not properly taken into account. 

In this work, we model the impact of baryonic processes on the cosmic shear via the baryonification emulator described in \S\ref{subsec:bcm}. We show the constraints on $S_8$ and $\Omega_{\rm m}$ that we obtain with and without scale cuts in the right panel of Fig.~\ref{fig:scale_cuts}. We obtain  $S_8=0.807^{+0.020}_{-0.022}$ with the DES scale cuts and $S_8=0.799^{+0.023}_{-0.015}$ without scale cuts. Interestingly, when applying the scale cuts, the marginalisation over baryons does not broaden significantly the $S_8$ constraints, although it slightly shifts the posterior towards high $S_8$, by $0.2\sigma$. This might be caused by a residual signal of baryonic effects in the data vector even after imposing the scale cuts, or it could also be a projection effect given by the unconstrained baryonic parameters. 

The marginalisation over 7 free baryonic parameters has significantly more impact when analysing all the angular scales: we find the constraint on $S_8$ $\approx 24\%$ weaker with respect to the gravity-only case, degrading part of the extra cosmological information contained in the small scales.   
However, we find no bias in the $S_8$-$\Omega_{\rm m}$ introduced by the addition of the small scales in the analysis. 
Moreover, we gain cosmological information when marginalising over baryons and going to smaller scales. When comparing the constraints we obtain with and without the angular scale cuts, we find $S_8$ and $\Omega_{\rm m}$ that are $10\%$ and $22\%$ tighter, respectively. 

This finding validates the robustness of the modelling of baryonic processes at all the scales employed in this work. 
We note that the marginalisation over 7 free baryonic parameters is a conservative choice, given that, as shown in App.~\ref{app:bcm_posteriors}, only one of these parameters is strongly constrained by our data. 

Therefore, adding extra information on baryonic processes, either with constraints from external datasets or educated guesses from hydrodynamical simulations, could help recover, at least partially, the cosmological information lost in the marginalisation. For instance, we could fix the unconstrained baryonic parameters to a value inferred with hydrodynamical simulations, analogously to \cite{Chen2023}. By setting all the parameters except $M_{\rm c}$ to the best-fitting parameters of the matter power spectrum of the hydrodynamical simulations BAHAMAS at $z=0$, we obtain $S_8=0.799^{+0.017}_{-0.017}$, in perfect agreement with our fiducial case, but $15\%$ tighter and with a better $\hat{\chi}^2$.

\begin{figure}
\includegraphics[width=0.45\textwidth]{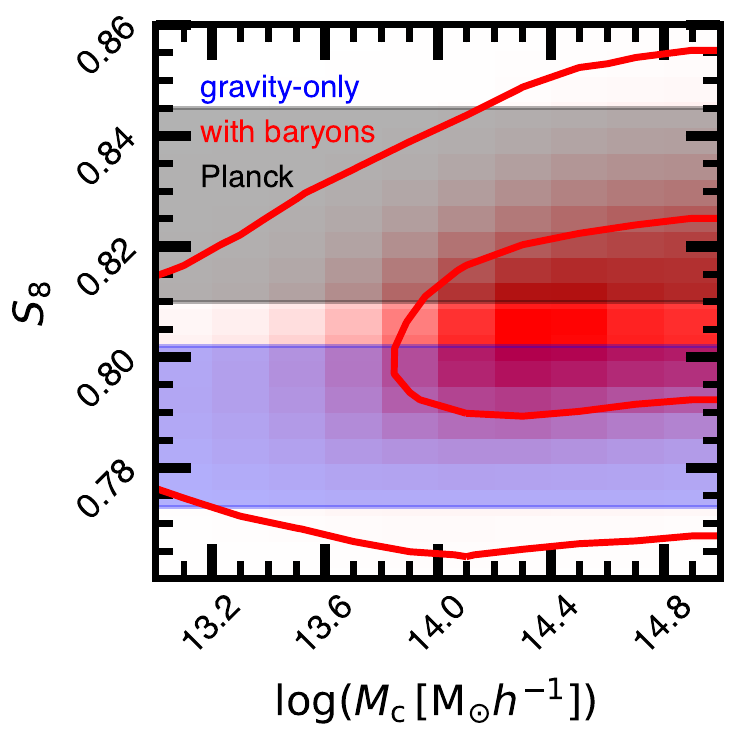}
\caption{Degeneracy between $S_8$ and $\log M_{\rm c}$, displaying the $1\sigma$ and $2\sigma$ credible levels (red), and using all angular scales. For comparison, we show as a horizontal band the $1\sigma$ constraints on $S_8$ obtained analysing all angular scales without modelling baryons (blue), and using the Planck TT+TE+EE+lowE posterior obtained by the DES Collaboration \protect\citep[black, ][]{Planck2018,Secco2022}.}
\label{fig:mc_degeneracy}
\end{figure}

\subsection{Constraints on baryons}
\label{subsec:baryons_constraints}
By including in our analysis all the angular scales available, we are able to constrain the astrophysical processes which modify the gravitational evolution of the cosmic density field. In particular, it has been shown that the suppression in the matter power spectrum is proportional to the gas fraction expelled from haloes by baryonic feedback processes \citep{Schneider&Teyssier2015,vandaalen2020}. We parametrise this fraction with $M_{\rm c}$, the characteristic halo mass for which half of the gas is depleted. 

\begin{figure}
\includegraphics[width=0.45\textwidth]{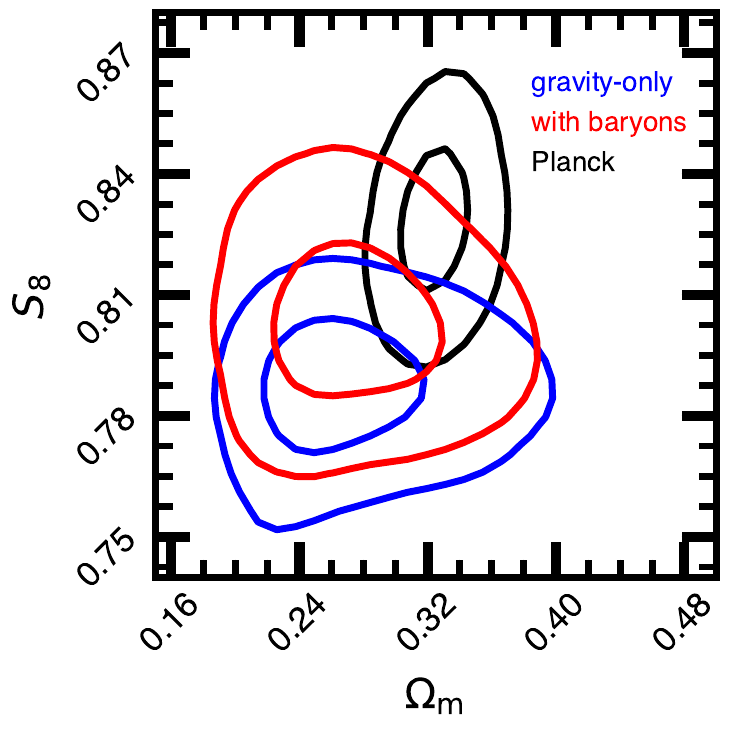}
\caption{$1\sigma$ and $2\sigma$ credible levels on $S_8$ and $\Omega_{\rm m}$ using the Planck TT+TE+EE+lowE posteriors of the DES Collaboration \protect\citep[black, ][]{Planck2018,Secco2022} and DES Y3 cosmic shear data, analysing all the angular scales either with our fiducial model (red) or without modelling baryons (blue), obtained in this work.}
\label{fig:s8_tension}
\end{figure}

In Fig.~\ref{fig:Mc} we show our constraints on $\log M_{\rm c}$ when including or not the small scales removed in the official DES analysis. 
When we apply the DES scale cuts, as expected, $\log M_{\rm c}$ is unconstrained, even if the data might have a residual sensitivity to baryonic effects, with higher values of $M_{\rm c}$ slightly preferred (blue line). 
When we analyse all the angular scales, we obtain a tight constraint $\log M_{\rm c} = 14.38^{+0.60}_{-0.56} \, [\log (\Msun)]$ (red line). 

 We find an excellent agreement between our estimate of $M_c$ and that obtained by \cite{Chen2023} (green line) which employed the same model as ours. However, our constraints are slightly weaker due to the different assumptions in the two analyses: first, \cite{Chen2023} employ an informative prior on cosmology, with all cosmological parameters fixed except for $\sigma_8$ and $\Omega_{\rm m}$, given by the 3x2pt analysis of DES Y3.
Second, \cite{Chen2023} used only the data points with angles smaller than the DES scale cuts, a TATT intrinsic alignment model, and they fixed all the baryonic parameters except $M_{\rm c}$ to the best-fitting values of the hydrodynamical simulation OWLS-AGN \citep{Schaye2010}. The different priors in cosmology (i.e. the extra information on cosmology retrieved by galaxy clustering, galaxy-galaxy lensing, and shear ratios) have likely the largest impact.

We observe that the posterior of $\log M_{\rm c}$ hits the boundary of its prior, at $\log M_{\rm c} = 15 \, [\log (\Msun)]$. This prior has been chosen to broadly encompass the current measurement of gas fractions in galaxy clusters measured in $X$-ray \citep{Vikhlinin2006,Arnaud2007,Sun2009,Giodini2009,Gonzalez2013}, as well as the prediction of hydrodynamical simulations \cite{McCarthy2017,McCarthy2018}. 
Therefore, we note that we are explicitly adding to our analysis prior information on the quantity of gas inside haloes, which cannot be lower than half the cosmic baryon fraction for haloes with mass $M_{\rm 200,c}=10^{15} \Msun$.  
Even if we argue that the prior on $M_{\rm c}$ is broad enough given X-ray data, we plan to build the next version of the baryonic emulator extending the prior $M_{\rm c}$ to higher values, to better cover the parameter space allowed by WL-only data. \\
The remaining 6 free baryonic parameters are unconstrained, and we show their posteriors in App.~{\ref{app:bcm_posteriors}}. \\
Finally, in Fig.~\ref{fig:Sk} we show the estimated suppression in the matter power spectrum at $z=0$. Specifically, we show the $1\sigma$ credible interval obtained taking into account the full posterior of the free cosmological and baryonic parameters in our analysis. 
We compare it to the BCM best-fitting models to several state-of-the-art hydrodynamical simulations, obtained in \cite{Arico2020c}: EAGLE \citep{Schaye2015,Crain2015,McAlpine2016}, Illustris \citep{Illustris2014}, Illustris TNG \citep{Pillepich2018,Springel2018}, OWLS-AGN \citep{Schaye2010}, and BAHAMAS \citep{McCarthy2017,McCarthy2018}. 
We infer a suppression of about $10\%$ at $k=2\ihMpc$, in broad agreement with the BAHAMAS suite and OWLS-AGN, but stronger than EAGLE and Illustris TNG and milder than Illustris. 
This finding is in perfect agreement with \cite{Chen2023}, although since we let free all the baryonic parameters, our model is more flexible e.g. at large scales (dominated by $\eta$, i.e. the distance range of the AGN feedback) and small scales (modulated by $M_{\rm z0,cen}$ and $\theta_{\rm inn}$, which regulate the galaxy-halo mass relation and inner shape of the gas, respectively).

\subsection{Correlation between cosmological and baryonic parameters}
\label{subsec:cosmo_bar_corr}
Constraints on cosmology and baryonic physics are not independent. For instance, the impact of AGN on the matter field depends on the amount of gas affected by feedback from supermassive black holes, and therefore on the cosmic baryon fraction $\Omega_{\rm b}/\Omega_{\rm m}$. Also, the amplitude of the linear density fluctuations, $\sigma_8$, affects in a minor and less trivial way the baryonic feedback \citep{Schneider2020,Arico2020c}. We can thus expect to find a correlation between $S_8$ and $M_{\rm c}$. We show the $1\sigma$ and $2\sigma$ credible levels of these two parameters in Fig.~\ref{fig:mc_degeneracy}. Indeed, we observe a weak but clear degeneracy, so that at high values of $M_{\rm c}$ correspond high values of $S_8$. For comparison, we display as bands the Planck TT+TE+EE+lowE $1\sigma$ intervals on $S_8$ obtained by the DES Collaboration \citep{Planck2018,Secco2022}, and our gravity-only analysis with all the angular scales. Following the degeneracy direction in $S_8-M_{\rm c}$, we can see how lower values of $M_{\rm c}$ agree with the $S_8$ values obtained by our DES gravity-only analysis, whereas going towards high values of $M_{\rm c}$ the value of $S_8$ becomes more compatible with Planck. We study the implications in the context of the so-called \say{$S_8$ tension} in \S\ref{sec:S8_tension}, but before, we discuss the constraints on intrinsic alignments from our small-scales analysis.

\subsection{Constraints on intrinsic alignment}
\label{subsec:ia_constraints} 

Our model allows us to constrain the intrinsic alignment of galaxies taking advantage of all the angular scales of DES Y3. With the NLA model and using DES angular scale cuts, we find $a_1=0.22^{+0.48}_{-0.30}$ and $\eta_1=4.11^{+0.86}_{-3.50}$. Interestingly, including small scales results in a tighter constrain on $\eta_1$ but not on $a_1$: removing scale cuts we infer $a_1=0.09^{+0.58}_{-0.25}$ and $\eta_1=4.50^{+0.48}_{-3.35}$. 

The NLA model seems to be sufficient to describe the full range of angular scales in DES. In fact, NLA is statistically preferred over TATT: we find a ratio of the Bayesian evidence $\mathcal{R} = \mathcal{Z}_{\rm NLA}/\mathcal{Z}_{\rm TATT}=3.0 \pm 1.2$. This value indicates a moderate/substantial preference for NLA over TATT, according to the commonly used (and somewhat arbitrary) Jeffreys scale \citep{Jeffreys1935}. 
For comparison, we find that the baryonic model with 7 free parameters is preferred to the gravity-only with a $\mathcal{R} = 1.5\pm 1.2$, and freeing only the baryonic parameter $M_{\rm c}$, $\mathcal{R} = 5.14\pm 1.2$. 
Despite this, the goodness of the fit is marginally better with TATT, $\hat{\chi}^2=1.033$, with respect to NLA $\hat{\chi}^2=1.043$.

Our inferred amplitudes of TATT in the DES galaxy sample are very low and consistent with zero, in agreement with \cite[][]{Secco2022,Amon2022}. That means that the intrinsic alignment contribution is subdominant with respect to the cosmological signal, and that simpler models are statistically preferred. 
We find also internal degeneracies in TATT which produce a multi-modal posterior of the tidal torque amplitude. We further analyse and discuss the results obtained with TATT in App.~\ref{app:ia_info}.  

\section{The $S_8$ tension}
\label{sec:S8_tension}

\begin{table*}
    \begin{tabular}{c c c c c }
    Model & \multicolumn{2}{c}{$S_8$} & \multicolumn{2}{c}{$\Omega_{\rm m}$}\\ \hline

    This work & DES scale cuts & all scales  & DES scale cuts & all scales  \\ \hline
    NLA BCM7 (fiducial) & $0.807^{+0.020}_{-0.022}$ & $0.799^{+0.023}_{-0.015}$ & $0.259^{+0.083}_{-0.040}$ & $0.252^{+0.066}_{-0.030}$\\
    NLA GrO & $0.802^{+0.019}_{-0.021}$ & $0.785^{+0.017}_{-0.012}$ & $0.253^{+0.084}_{-0.030}$ & $0.249^{+0.061}_{-0.035}$\\
    NLA BCM1 & $0.803^{+0.019}_{-0.020}$ & $0.799^{+0.017}_{-0.017}$ & $0.262^{+0.073}_{-0.044}$ & $0.253^{+0.066}_{-0.036}$\\
    NLA BCM-extreme & $0.838^{+0.030}_{-0.022}$ & $0.829^{+0.023}_{-0.022}$ & $0.243^{+0.094}_{-0.025}$ & $0.272^{+0.078}_{-0.036}$\\ \hline
    
    TATT BCM7 & $0.796^{+0.022}_{-0.028}$ & $0.788^{+0.027}_{-0.035}$ & $0.258^{+0.084}_{-0.031}$ & $0.227^{+0.065}_{-0.029}$\\
    TATT GrO & $0.794^{+0.021}_{-0.030}$ & $0.755^{+0.028}_{-0.023}$ & $0.248^{+0.076}_{-0.029}$ & $0.253^{+0.057}_{-0.039}$\\
    TATT BCM1 & $0.791^{+0.027}_{-0.025}$ & $0.790^{+0.022}_{-0.035}$ & $0.252^{+0.073}_{-0.041}$ & $0.235^{+0.052}_{-0.030}$\\
    TATT BCM-extreme & $0.828^{+0.029}_{-0.032}$ & $0.816^{+0.034}_{-0.030}$ & $0.267^{+0.065}_{-0.052}$ & $0.237^{+0.060}_{-0.031}$\\ \hline 
    DES Collaboration &  &  &  &  \\ \hline
    DES NLA & $0.783^{+0.024}_{-0.022}$ & - & $0.289^{+0.080}_{-0.045}$ & -\\
    DES NLA + SR & $0.775^{+0.017}_{-0.022}$ & - & $0.291^{+0.042}_{-0.055}$& - \\
    DES TATT & $0.767^{+0.030}_{-0.039}$ & - & $0.270^{+0.071}_{-0.042}$ & -\\
    DES TATT + SR (fiducial) & $0.763^{+0.021}_{-0.026}$ & - & $0.269^{+0.059}_{-0.044}$ & -\\ \hline 
    \end{tabular}
     \centering
    \caption{Constraints (mode of the 1D marginalised posteriors and respective 34th percentiles) on $S_8$ and $\Omega_{\rm m}$ obtained applying DES scale cuts and using all the angular scales. We explore different models, with NLA or TATT intrinsic alignment, gravity-only, or applying a baryonification with 1 or 7 free parameters (BCM1, BCM7, respectively). We also include a model with an extreme implementation of AGN feedback. We report for comparison the constraints we get analysing the DES Collaboration chains, with and without shear ratios (SR).}
\label{tab:model_params}
\end{table*}

\begin{figure}
\includegraphics[width=0.49\textwidth]{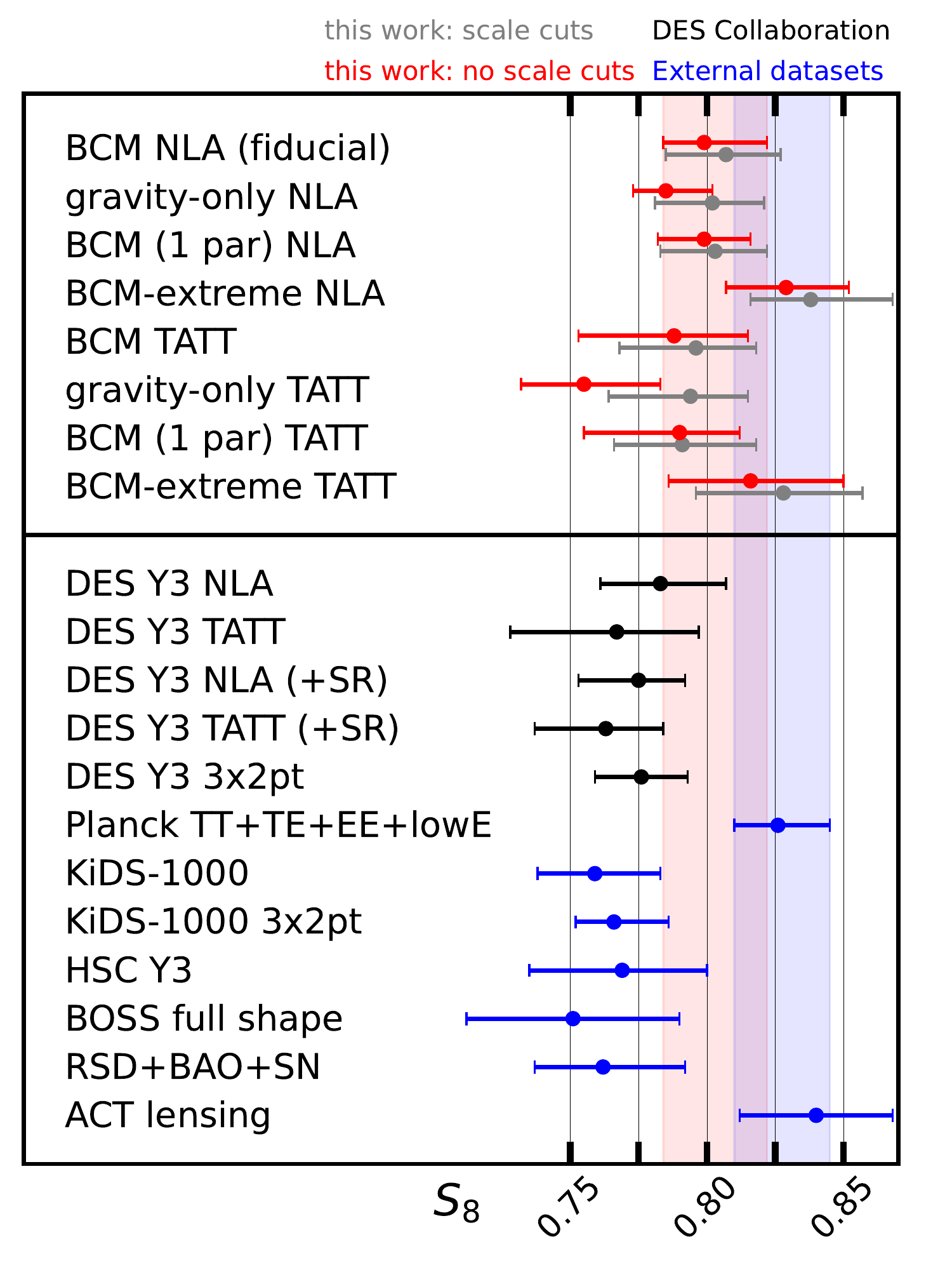}
\caption{
$S_8$ constraints obtained in this work with different analysis strategies of DES Y3 data, either applying the scale cuts (grey data points) or employing all the angular scales (red data points). We highlight our fiducial model by displaying a red band correspondent with its $1\sigma$ credible interval.
We employ our baryonification (BCM) with 1 or 7 free parameters, or fix all the parameters to extreme baryonic feedback. This last model, dubbed {\it BCM-extreme}, is in tension with X-ray observations, and should be considered as an upper limit for baryonic effects.  
We also vary the intrinsic alignment model choosing either NLA or TATT. For comparison, in the lower part of the panel we display the constraints obtained from the DES Collaboration chains with scale cuts, adding or not the shear ratios \protect\citep[black data points, ][]{Secco2022,Amon2022,DESY3_3x2pt}. Moreover, we include the external datasets specified in the legend \protect\citep[blue data points, ][]{Planck2018,Asgari2021,Li2023,Philcox&Ivanov2022,Nunes&Vagnozzi2021,Madhavacheril2023}, which we describe in more detail in the main text. We highlight with a blue band the $1\sigma$ region preferred by Planck.
}
\label{fig:all_constraints}
\end{figure}

In this section, we compare the cosmological constraints that we have obtained analysing the cosmic shear of DES Y3 against external datasets, in light of the so-called \say{$S_8$ tension}. We also compare our results with the official ones obtained by the DES Collaboration, and discuss the main differences in the analyses.   

Throughout this section, we will quantify the tension among datasets or analyses by approximating the marginalised posteriors as Gaussian-distributed functions, and considering the mean and standard deviation following e.g. \cite{Heymans2021} \citep[for a method that takes into account non-Gassianity, see e.g.][]{Raveri&Doux2021}.

\subsection{Impact of different model assumptions on the $S_8$ tension}

In Fig.~\ref{fig:s8_tension} we compare our results against results from the analysis of Planck's satellite data including temperature and polarization measurements (TT+TE+EE+lowE power spectra). Specifically, we use the chains made available by the DES Collaboration\footnote{\url{https://desdr-server.ncsa.illinois.edu/despublic/y3a2_files/chains/}}, where the Planck likelihood is sampled over the same prior-space used in the DES analysis (and therefore almost identical to ours). We also show our fiducial analysis (no scale cuts, explicit model for baryons, NLA for intrinsic alignments, and emulators for the matter power spectrum) as red lines, and the gravity-only case as blue lines. 

We can see that our marginalised posteriors on $S_8$ are in $1.9\sigma$ tension in the gravity-only case compared to Planck. However, the tension reduces to $0.9\sigma$ when marginalising over baryonic effects. Therefore, our data suggest that at present Planck and DES Y3 are not statistically in tension. Additionally, the agreement between the data could potentially increase further if, for instance, external datasets (e.g. X-ray gas fraction) constrain the baryonic parameters to relatively strong feedback.  

Combining such datasets is a non-trivial task, due to different systematics, e.g. hydrostatic mass bias and data covariances, and it is outside the scope of this work. However, this could be an interesting avenue to investigate in the future.

To explore further how the $S_8$ posterior is affected by the modelling of baryons, we consider two different scenarios. First, we fix all the baryonic parameters except for $M_{\rm c}$ to the best-fitting values of the matter power spectrum measured in the hydrodynamical simulation BAHAMAS at $z=0$.\footnote{ These values, obtained by \cite{Arico2020c}, are $\log{\eta_{b}}=-0.33$, $\log{\beta_{b}}=-0.28$, $\log M_{z0,\rm cen}=10.21 [\log (\Msun)]$, $\log{\theta_{\rm inn}}=-0.62$, $\log{\theta_{\rm out}}=0.12$, $\log{M_{\rm inn}}=9.95 [\log (\Msun)]$.} Second, to have a sense of what is the most extreme shift in $S_8$ that baryonic processes can cause, we set the baryonic parameters to the values that maximise the feedback allowed by our emulator.\footnote{We thus set $\log{M_{\rm c}}= 15 [\log (\Msun)]$, $\log{\eta_{b}}=-0.7$, $\log{\beta_{b}}=0.7$, $\log M_{z0,\rm cen}= 9 [\log (\Msun)]$, $\log{\theta_{\rm inn}}=-0.53$, $\log{\theta_{\rm out}}=0$, $\log{M_{\rm inn}}=9 [\log (\Msun)]$.} We note that these parameters are already ruled out by X-ray data, even if we find that they still provide a good fit to the DES Y3 cosmic shear data. We dubbed this model {\it BCM-extreme}. 

We summarise all our $S_8$ constraints in Fig.~\ref{fig:all_constraints} and in Tab.~\ref{tab:model_params}. We note that the $S_8$ value inferred with and without scale cuts are generally in good mutual agreement, except for the gravity-only models where they present $0.5\sigma$ (with NLA) and $0.9\sigma$ (with TATT) shifts. 
In particular, our gravity-only analysis with DES scale cuts differs by less than $1\sigma$ from Planck, when using either NLA or TATT. We can conclude that DES angular scale cuts are reliably removing the baryonic effects, which must be accounted for only when analysing the full DES data vector. This is not true for scenarios with very high baryonic feedback. For example, in our {\it BCM-extreme} case, we find an excellent agreement between the DES Y3 cosmic shear and Planck (tension of $0.1\sigma$), both with and without scale cuts. However, we stress that this a very unlikely baryonic scenario, where all the gas within haloes up to $M=10^{15} \Msun$ has been pushed for tens of Mpc, and must be simply taken as an extreme upper limit of the impact of baryonic processes.

Finally, by comparing the $S_8$ values obtained by varying intrinsic alignment models, we observe in Fig.~\ref{fig:all_constraints} that the $S_8$ posteriors are generally broader and shifted towards lower values when using TATT. This trend could be caused by internal degeneracies and projection effects of the TATT parameters, which are allowed to vary over a broad parameter space that is not physically motivated, as speculated also by \cite{Secco2022}. 
We explore this in more detail in App.~\ref{app:ia_info}. 

\begin{figure}
\includegraphics[width=0.4\textwidth]{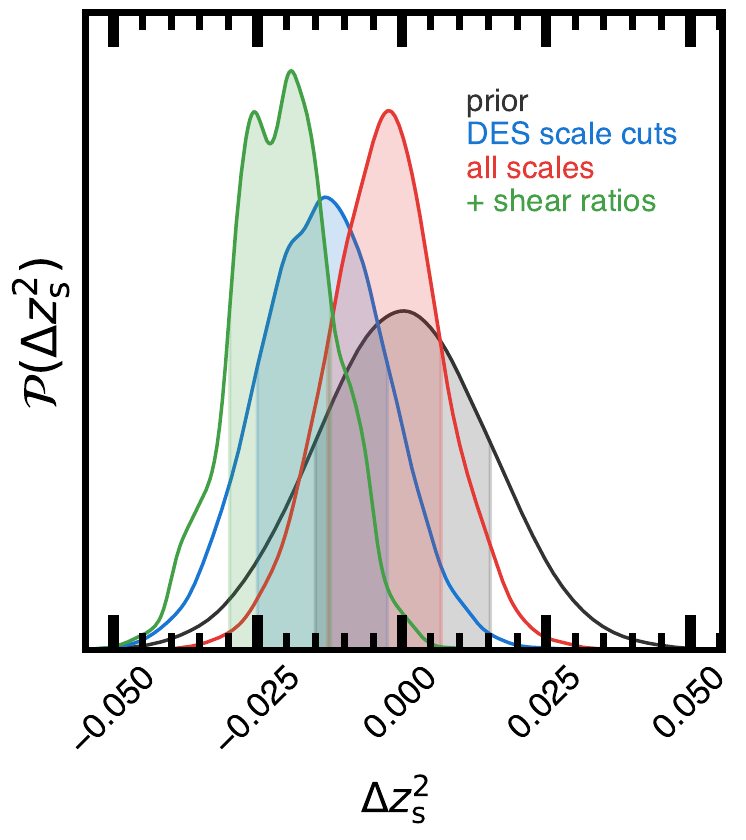}
\caption{Prior (black) and posteriors of the photo-z shift in the second tomographic bin $\Delta z_{\rm s}^2$, when using DES scale cuts (blue) and all angular scales (red). For comparison, we show the posteriors from the DES Collaboration chains, with scale cuts and the addition of the lensing shear ratios \protect\citep[green,][]{Amon2022,Secco2022}.}
\label{fig:deltaz}
\end{figure}

\subsection{Comparison with the official DES Y3 analysis}

Our constraint on $S_8$ is systematically higher than that obtained with the same dataset by the DES Collaboration \citep{Secco2022,Amon2022}. Even when applying the DES scale cuts, our gravity-only results are $1.4\sigma$ away from those of the DES Collaboration. This is a significant discrepancy since both analyses use exactly the same shear correlation functions. We now explore the origin of this discrepancy.

\begin{figure}
\includegraphics[width=0.45\textwidth]{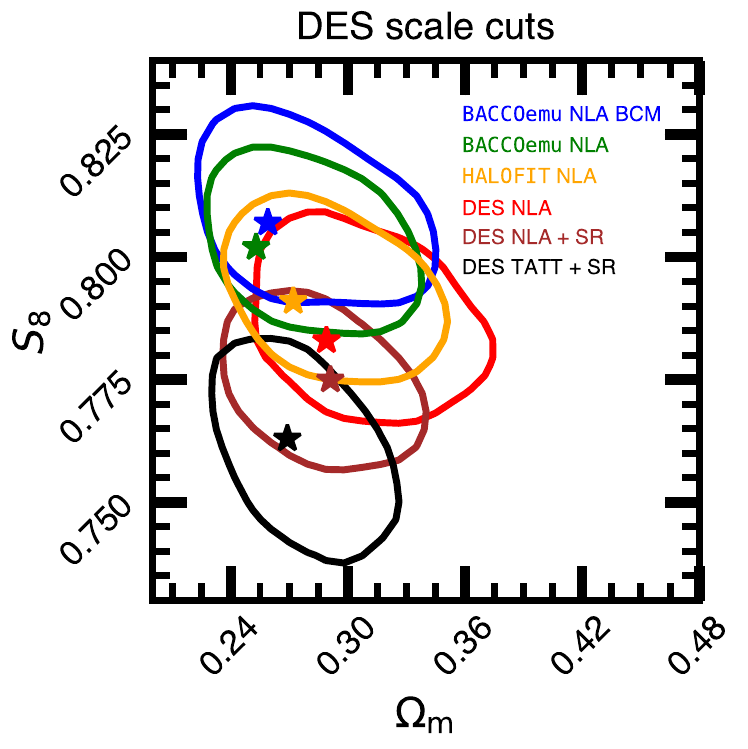}
\caption{Impact of the different modelling choices to explain the different fiducial cosmological constraints obtained by the DES Collaboration \protect\citep[DES TATT + SR,][]{Secco2022,Amon2022} and this work ({\code BACCOemu} NLA BCM). We show the $1\sigma$ credible level of $S_8$ and $\Omega_{\rm m}$ (solid lines) and maximum posteriors (stars), obtained by the DES Collaboration with the {\code CosmoSIS} pipeline and different choices regarding the intrinsic alignment model and the use of shear ratios (black, brown, red, according to the legend). Analogously, we show also our results, obtained with the {\code BACCO} pipeline using {\code HALOFIT} (orange), or {\code BACCOemu} without and with baryons (green and blue, respectively).       
}
\label{fig:s8_transition}
\end{figure}

In Fig.~\ref{fig:s8_transition} we illustrate the impact of various modelling ingredients and choices in the $S_8-\Omega_m$ constraints. Lines show the $1\sigma$ credibility contours whereas stars highlight the mode of the marginalised posteriors. The official DES and our fiducial results are shown in black and blue, respectively. We now discuss specific differences in the analyses.

\textit{Intrinsic alignments:} An important difference is the choice of the fiducial intrinsic alignment model. We estimate a shift toward high $S_8$ between $0.4\sigma$ and $0.7\sigma$ when using NLA instead of TATT, depending on modelling choices e.g. scale cuts, baryonic modelling, and shear ratios. As we have argued before (\S\ref{subsec:ia_constraints}), although TATT is in principle a more complete description of intrinsic alignments, the additional complexity is not justified by the current data. This is the case for the DES analysis and scale cuts as well as for our approach including small scales.

\textit{Shear ratios:} Another difference is that we do not employ the lensing shear ratios. In the DES analysis, excluding shear ratios increases $S_8$ by $\sim0.1$ and $0.3\sigma$ when employing NLA and TATT, respectively. The shift is arguably caused by stronger constraints on the photo-z uncertainties and intrinsic alignment parameters. 
In agreement with our finding, \cite{Secco2022,Amon2022} report that the inclusion of shear ratios shifts the intrinsic alignment amplitude towards slightly negative values (although still compatible with zero). This is not expected physically in the absence of systematic errors in the data. On the other hand, with a physically-motivated prior $a_1 \in [0,5]$, these authors find that their posteriors on $S_8$ do not shift significantly. However, the impact is dependent on the intrinsic alignment model used and the addition or not of shear ratios.

\textit{Photo-z distributions:} An advantage of using shear ratios is that they provide information on the redshift distribution of background galaxies. Notably, we find that small scales provide a similar level of information on photometric redshifts. It is thus interesting to compare their constraints since including shear ratios or not does affect the inferred $S_8$ value. We show, as an example, the posterior of $\Delta z_{\rm s}^2$ in Fig.~\ref{fig:deltaz} (see Appendix~\ref{sec:deltaz_posteriors} for other redshift bins). This is the photo-z parameter that differs the most between our analysis and that of DES with shear ratios.

Firstly, we see that including the small scales provides a constraining power on $\Delta z_{\rm s}^2$ competitive with that of shear ratios, even after marginalisation over baryons. Specifically, in our fiducial analysis we obtain $\sigma[\Delta z_{\rm s}^2]\sim0.012$ when applying DES scale cuts, and $\sim0.01$ when including small scales -- a comparable precision to that by including shear ratios $\sim0.009$. Of course, our approach has the advantage of not relying on the modelling of the galaxy-galaxy lensing, and thus galaxy bias, lenses redshift distributions, etc. 

 Moreover, we note that our posterior on $\Delta z_{\rm s}^2$ is in good agreement with the Gaussian prior and with the analysis with DES scale cuts. 
 Instead, when using shear ratios, the data prefer a shift in the mean photo-z of $\sim0.0025$. Given the sensitivity of the shear signal to the sources' redshift distributions, these might cause part of the small shift in the cosmological constraints observed when comparing results with and without shear ratios.
 The information contained in the small scales of the cosmic shear is not in alarming tension with the one contained in the small scales of shear ratios. However, as the precision of WL surveys increases, this comparison can provide a good sanity check to highlight possible shortcomings of the modelling of cosmic shear and galaxy-galaxy lensing.     


\textit{Nonlinear modelling:} Another difference is that we employ a more precise model for the nonlinear matter power spectrum, especially in the case of massive neutrinos. We show in Appendix~\ref{app:nl_info} the difference in our cosmology constraints obtained when using our non-linear emulator or {\code HALOFIT} used by the DES Collaboration. By using {\code BACCOemu}, the $S_8$ constraints are shifted by $0.4\sigma$ toward high values when applying DES scale cuts, and up to $0.5$ when considering all angular scales. The slightly different choice for the $M_{\nu}$ prior has a comparatively small effect of $0.1\sigma$.

\textit{Baryonic modelling and pipeline:} Finally, we note an extra $0.2\sigma$ shift caused by the marginalisation over baryons when applying DES scale cuts. We should also consider a $0.1\sigma$ shift given by the pipeline implementations in {\code BACCO} and {\code CosmoSIS}. 
 
To summarise, we find that the fiducial modelling of intrinsic alignment and non-linearities cause the largest shift in $S_8$. Notably, most of the effects listed here shift the $S_8$ posterior toward higher values. Thus, as shown in Fig.\ref{fig:s8_transition}, also smaller effects (with DES scale cuts) e.g. the baryon marginalisation and the addition of shear ratios, other than the difference in $M_{\nu}$ prior, sum up to make the final $1.4\sigma$ discrepancy that we report.  

\subsection{Comparison with other datasets}

When assuming $\Lambda$CDM, several authors have claimed that LSS observations prefer statistically lower values of $S_8$ compared to temperature and polarization fluctuations measured by Planck \citep{Planck2018}. For instance, by using only BOSS data and exploiting the full shape of the galaxy power spectrum and bispectrum, \cite{Philcox&Ivanov2022} inferred $S_8=0.751 \pm 0.039$ -- a figure that is in agreement with lensing measurements and lower than Planck. \cite{Nunes&Vagnozzi2021} used a compilation of growth rates from the redshift space distortions (RSD) measured in different surveys, combined with Baryonic Acoustic Oscillations (BAO) and type Ia supernovae, and found $S_8=0.762^{+0.030}_{-0.025}$. 

Regarding weak lensing data, \cite{Secco2022,Amon2022} found that DES Y3 cosmic shear is at $2.3\sigma$ tension with Planck, according to the so-called Bayesian Suspiciousness \citep{Handley&Lemos2019}, where the prior volume effects are removed from the Bayes ratios. When projected to the $S_8$ parameters, they found $S_8=0.759^{+0.025}_{-0.023}$ while Planck data suggests $S_8=0.826^{+0.019}_{-0.016}$. Adding galaxy-galaxy lensing and galaxy clustering (3x2pt analysis) the Suspiciousness lowered down to $0.7\sigma$. 

Similarly, with the cosmic shear analysis of the Kilo Degree Survey (KiDS), \cite{Asgari2021} measured $S_8=0.759^{+0.024}_{.0.021}$ and reported a $3\sigma$ tension in the $S_8$ posterior. Combining the KiDS cosmic shear analysis with the redshift-space galaxy clustering from the Baryon Oscillation Spectroscopic Survey \citep[BOSS,][]{BOSS2015} and the 2-degree Field Lensing Survey \citep[2dFLenS,][]{Blake2016}, \cite{Heymans2021} got $S_8=0.766^{+0.020}_{-0.014}$, and claimed a tension with Planck between $2\sigma$ and $3\sigma$. 


Recently, the Subaru Hyper Supreme-Cam (HSC) Collaboration published the cosmic shear analysis after the third year of data collection, reporting  $S_8=0.769^{+0.031}_{-0.034}$ with correlation functions \citep{Li2023}, and $S_8=0.776^{+0.032}_{-0.033}$ with power spectra \citep{Dalal2023}.

Our fiducial constraint on $S_8$ is systematically higher but in broad agreement with results from the shear-only analyses of these surveys. Our preferred value is approximately $1.5\sigma$ higher than that of KIDS-1000, $1.4\sigma$ higher than in DES Y3, and $0.8\sigma$ above HSC Y3 (analysis with correlation functions). It is, however, unclear how much of this tension would be reduced by homogenising analysis choices and the improvements we adopt in our pipeline, given the different scales and redshifts probed in the analyses. At the small scales, KiDS-1000 stops at 0.5 arcmins in $xi_+$ and 4 arcmins in $xi_-$, whereas HSC Y3 to $7.1$ arcmins in $\xi_+$ and 31.2 arcmins in $\xi_-$. Both the surveys model the baryonic suppression with {\code HMcode} \citep{Mead2015}. 

Using more refined baryonic modelling, the BCM emulator described in \cite{Giri&Schneider2021},  \cite{Schneider2022} analysed the cosmic shear measured by KiDS-1000 using external data from X-ray and kinetic Sunyaev-Zel'dovich (kSZ).
In agreement with our results, they found that baryons reduce the tension with Planck, from $3.8\sigma$ to $2.6\sigma$ in KiDS. 
They notice that KiDS-1000 data alone are not enough to constrain baryonic physics (whereas we find that DES Y3, arguably due to the larger sky area, can constrain the most important parameter, $M_{\rm c}$). 
By adding X-ray and kinetic Sunyaev-Zel'dovich (SZ) data, they were able to constrain 3 out of 7 baryonic parameters, and measure a mild-strong feedback broadly in agreement with that measured with DES Y3. 
Overall, they find a similar impact of baryons on cosmology to what we find when analysing all the angular scales of DES Y3, a shift in $S_8$ of about $1\sigma$ towards higher $S_8$. 

Using HMcode \citep{Mead2020b} as a model for baryons, \cite{Troster2022} combined the cosmic shear from KiDS-1000 with the cross-correlation between shear and thermal SZ (tSZ), measured by Planck and the Atacama Cosmology Telescope (ACT) \citep{MallabyKay2021}. They find an improvement on the $S_8$ constraint of $40\%$ in the joint analysis but did not reduce the S8 tension: they infer $S_8=0.751^{+0.020}_{-0.017}$, at $3.4\sigma$ from Planck (see also a similar analysis by \cite{Robertson2021} cross-correlating KiDS with Planck/ACT CMB lensing. However, since HMcode has been explicitly calibrated to reproduce the BAHAMAS hydrodynamical simulations, it is not clear whether these results would hold with more flexible baryonic modelling such as that we propose here. 

Interestingly, a very recent analysis of the CMB lensing from ACT DR6, in combination with BAO data, reports $S_8=0.849 \pm 0.028 $ \citep{Madhavacheril2023,Qu2023}, in perfect agreement with the inference from Planck CMB anisotropies.

A different way to improve the accuracy of current constraints is to take advantage of the fact that, at the moment, LSS surveys are largely independent of each other. Combining 6 different cosmic shear, galaxy clustering, and CMB lensing surveys, \cite{Garcia2021} placed a tight constraint $S_8=778^{+0.009}_{-0.009}$, $3.4\sigma$ in tension with Planck. 
Unfortunately, precise non-linearities and baryons were not included in this analysis. Additionally, the modelling inevitably requires several assumptions regarding the way in which galaxies trace the underlying matter field. 

Finally, as we have discussed in the previous section, our analysis pipeline and choices deliver constraints that are somewhat in tension with those of the official DES Collaboration. Thus, it would be interesting in the future to re-analise jointly the data from KiDS and HSC including small-scale information and the advances we have described in this paper. Furthermore, as these WL surveys provide measurements that are at different redshifts, they could be combined to further constrain the time evolution of baryonic processes. 

Similarly, it is important to study more in-depth the role of the intrinsic alignment models and photo-z systematic errors. Finally, to robustly combine WL with clustering data, it will be essential to carefully consider the accuracy of current models for galaxy biasing. For instance, correlations between galaxy number and halo properties, baryons, and assembly bias all affect galaxy clustering and galaxy-galaxy lensing in different manners \citep[see e.g.][]{Chaves2023,Contreras2023}, but are commonly neglected in mock catalogues used to validate lensing pipelines. We plan to address all these issues in the future.

\section{Conclusions}
\label{sec:conclusions}
In this paper we have analysed the cosmic shear correlation functions measured in DES Y3, exploiting for the first time all the angular scales measured. We have implemented a new fast pipeline to predict the cosmic shear correlation functions, and employed the neural network emulators of {\code BACCOemu} to accurately predict the matter power spectrum with non-linearities and baryonic effects. 
Our main findings are the following:
\begin{itemize}

    \item We find additional information at small scales. Constraints on $S_8$ are tighter by $30\%$ when ignoring astrophysical processes, and still $10\%$ ($15\%$) tighter when marginalising over 7 (1) baryonic parameters. These numbers are specific to DES and we expect the gains to be larger in upcoming surveys.
    
    \item The $S_8$ posterior shifts by $\approx 0.5\sigma$ towards lower values when removing DES scale cuts and not modelling baryonic effects. The cosmological inference is instead robust against scale cuts when modelling baryons with a baryonification algorithm (Fig.~\ref{fig:scale_cuts});

    \item We constrain baryonic feedback via the BCM parameter $M_{\rm c}$. We obtain $\log M_{\rm c} [\Msun] = 14.38^{+0.60}_{-0.56}$ which means haloes with mass $M_{200,c}=10^{14} \Msun$ have lost half of their gas reservoir (Fig.~\ref{fig:Mc}). 
    
    \item We find a correlation between $S_8$ and $M_{\rm c}$, so that stronger baryonic effects correspond to higher $S_8$. In light of the \say{$S_8$} tension, this means that stronger baryonic feedback increases the agreement between DES Y3 and Planck (Fig.~\ref{fig:mc_degeneracy}). 
    
    \item We infer $S_8 = 0.799^{+0.023}_{-0.015}$ and $\Omega_{\rm m}=0.252^{+0.066}_{-0.030}$ when including the entire range of scales in DES and conservatively marginalising over 7 free baryonic parameters. This value $S_8$ is $0.9\sigma$ lower than that preferred by Planck TT+TE+EE+lowE data (Fig.~\ref{fig:s8_tension}).
    
    \item Our inferred value of $S_8$ differs from that inferred in the official DES analysis by $1.4\sigma$ (Fig.~\ref{fig:s8_transition}). This is a significant difference since both analyses employ very similar datasets. We attribute the discrepancy to five factors: i) the modelling of intrinsic alignments; ii) the computation of the nonlinear power spectrum; iii) the modelling of baryons; iv) the employment of shear ratios by the DES Collaboration; v) minor difference between the pipeline and priors used.
\end{itemize}

We can conclude that, with current data, the modelling details impact significantly the parameter inference, and therefore the assessment of tensions between different datasets. Improving the modelling, as well as explicitly taking into account the theoretical uncertainties in the analysis, appears to be a key step to delivering robust cosmological constraints.

We have shown that we can accurately model the cosmic shear of DES Y3, down to the smallest scales available. At these scales, cosmology and astrophysical processes are tightly intertwined, and disentangling them is not an easy task. 

In the future, we plan to do it by consistently joining different datasets. Specifically, we will perform a 3x2pt analysis, modelling also galaxy-galaxy lensing and galaxy clustering to very small scales, by exploiting non-linear bias emulators included in {\code BACCOemu} \citep{Zennaro2021,Pellejero2022,Pellejero2022b}. To inform the baryonic model, we plan to use X-ray, thermal \& kinetic Sunyaev-Zel'dovich datasets, and their cross-correlations when possible. Such an analysis will be of particular importance in light of the upcoming surveys, such as Euclid and LSST, where the solidity of our modelling framework will be stress-tested in an unprecedented way. 

\section*{Acknowledgements}
We thank David Alonso, Dragan Huterer, Marco Gatti, Francisco Maion, Aurel Schneider, Lucas Secco, and the anonymous referee, for carefully reading the manuscript and providing valuable feedback. We additionally thank Aurel Schneider and Lucas Secco for sharing the chains and results of their analyses, and Marco Gatti for inspiring Fig.~\ref{fig:s8_transition}. We thank all the members of the cosmology group at DIPC for stimulating discussions and help in running $N$-body simulations. The authors acknowledge the support of the E.R.C. grant 716151 (BACCO). REA acknowledges the support of the Project of Excellence Prometeo/2020/085 from the Conselleria d'Innovació, Universitats, Ciéncia i Societat Digital de la Generalitat Valenciana, and of the project PID2021-128338NB-I00 from the Spanish Ministry of Science. SC acknowledges the support of the ``Juan de la Cierva Incorporac\'ion'' fellowship (IJC2020-045705-I). C.H.-M. acknowledges the support of project PID2021-126616NB-I00 from the Spanish Ministry of Science. The authors also acknowledge the computer resources at MareNostrum and the technical support provided by Barcelona Supercomputing Center (RES-AECT-2019-2-0012 \& RES-AECT-2020-3-0014). We acknowledge the use of the following software:  {\code BACCO} \& {\code BACCOemu} \citep{Angulo2020,Arico2020c,Arico2021}, {\code Fast-PT} \citep{McEwen2016,Fang2017}, {\code POLYCHORD} \citep{Handley2015}, {\code CosmoSIS} \citep{Zuntz2015}, Core Cosmology Library \citep[{\code CCL}, ][]{CCL2019}, {\code NumPy} \citep{numpy}, {\code mcfit}\footnote{\url{https://github.com/eelregit/mcfit}}, {\code SciPy} \citep{scipy}, {\code Matplotlib} \citep{matplotlib}, {\code ChainConsumer} \citep{Hinton2016}, {\code corner} \citep{corner}, {\code anesthetic} \citep{Handley2019}. 
The data underlying this article will be shared on reasonable request to the corresponding author.
This project used public archival data from the Dark Energy Survey (DES). Funding for the DES Projects has been provided by the U.S. Department of Energy, the U.S. National Science Foundation, the Ministry of Science and Education of Spain, the Science and Technology FacilitiesCouncil of the United Kingdom, the Higher Education Funding Council for England, the National Center for Supercomputing Applications at the University of Illinois at Urbana-Champaign, the Kavli Institute of Cosmological Physics at the University of Chicago, the Center for Cosmology and Astro-Particle Physics at the Ohio State University, the Mitchell Institute for Fundamental Physics and Astronomy at Texas A\&M University, Financiadora de Estudos e Projetos, Funda{\c c}{\~a}o Carlos Chagas Filho de Amparo {\`a} Pesquisa do Estado do Rio de Janeiro, Conselho Nacional de Desenvolvimento Cient{\'i}fico e Tecnol{\'o}gico and the Minist{\'e}rio da Ci{\^e}ncia, Tecnologia e Inova{\c c}{\~a}o, the Deutsche Forschungsgemeinschaft, and the Collaborating Institutions in the Dark Energy Survey.
The Collaborating Institutions are Argonne National Laboratory, the University of California at Santa Cruz, the University of Cambridge, Centro de Investigaciones Energ{\'e}ticas, Medioambientales y Tecnol{\'o}gicas-Madrid, the University of Chicago, University College London, the DES-Brazil Consortium, the University of Edinburgh, the Eidgen{\"o}ssische Technische Hochschule (ETH) Z{\"u}rich,  Fermi National Accelerator Laboratory, the University of Illinois at Urbana-Champaign, the Institut de Ci{\`e}ncies de l'Espai (IEEC/CSIC), the Institut de F{\'i}sica d'Altes Energies, Lawrence Berkeley National Laboratory, the Ludwig-Maximilians Universit{\"a}t M{\"u}nchen and the associated Excellence Cluster Universe, the University of Michigan, the National Optical Astronomy Observatory, the University of Nottingham, The Ohio State University, the OzDES Membership Consortium, the University of Pennsylvania, the University of Portsmouth, SLAC National Accelerator Laboratory, Stanford University, the University of Sussex, and Texas A\&M University.
Based in part on observations at Cerro Tololo Inter-American Observatory, National Optical Astronomy Observatory, which is operated by the Association of Universities for Research in Astronomy (AURA) under a cooperative agreement with the National Science Foundation.

\bibliographystyle{mnras}
\bibliography{bibliography} 


\begin{appendix}
\section{Modelling of non-linearities}
\label{app:nl_info}

\begin{figure}
\includegraphics[width=0.4\textwidth]{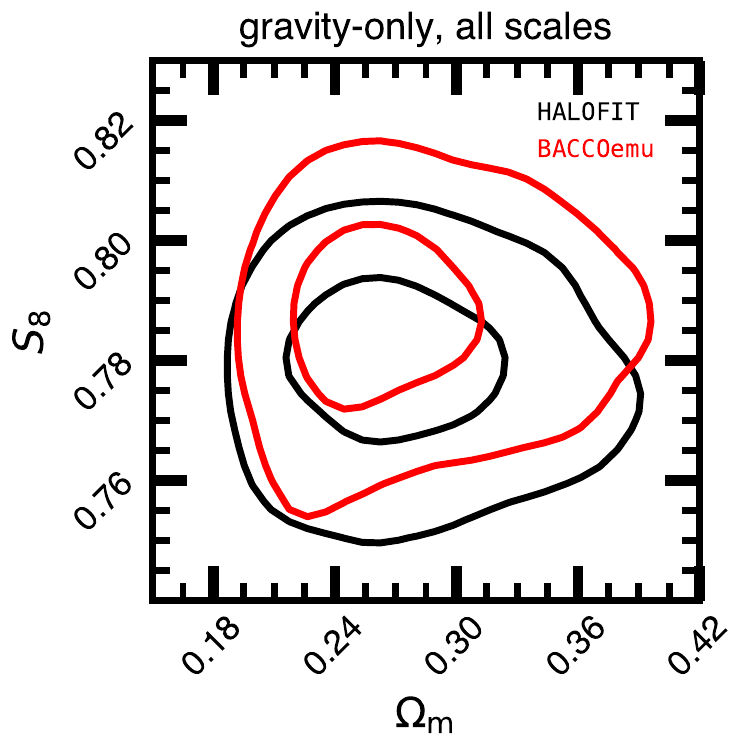}
\caption{$1\sigma$ and $2\sigma$ credible interval on $S_8$ and $\Omega_{\rm m}$ when modelling the matter power spectrum with {\code HALOFIT} (black) or {\code BACCOemu} (red). All the angular scales are used, not considering baryonic processes.}
\label{fig:halofit}
\end{figure}

The DES Collaboration employs {\code HALOFIT} \citep{Takahashi2012} to predict the nonlinear matter power spectrum \citep[e.g.][]{Secco2022,Amon2022}. {\code HALOFIT} has a nominal accuracy of $10\%$ at $k=10 \, \ihMpc$ \citep{Takahashi2012}. \cite{Krause2021} have shown that accuracy is enough to model the shear signal over the scales relevant for DES Y3, i.e. it causes a shift $<0.3\sigma$ compared to the  {\code EuclidEmulator} and {\code CosmoEmu} emulators \citep{EuclidEmulator2,Heitmann2014}. 

Generally, the accuracy of {\code HALOFIT}, as well as {\code HMcode}, {\code BACCOemu}, and other emulators, has been tested only over a relatively small cosmological space with respect to the prior used in DES analysis or even compared to the $\sim2-3\sigma$ regions of the posteriors. This could cause bias in cosmological parameters as the error could be significantly larger in some regions of the parameter space, for instance, for very massive neutrinos.

To test the performance of {\code HALOFIT} in a larger cosmological space, we have carried out a suite of 20 $N-$body simulations with cosmological parameters randomly distributed over the prior space. These simulations have a mass resolution so that the power spectrum is converged at a few per cent level at $z=0$ and $k=10\ihMpc$. We find that {\code HALOFIT} performs remarkably well -- considering that it was fitted to only 16 $N-$body simulations -- achieving a precision of around $20\%$. This, however, could still introduce significant biases in our cosmological constraints. 

Building emulators over large cosmological space volumes is very challenging. This would in principle require a large number of high-resolution simulations. Alternatively, here we exploit the  \say{cosmology scaling} of \cite{A&W2010,Angulo2020}. Therefore, we build an updated version of {\code BACCOemu} by employing a suite of 35 $N-$body simulations run with various cosmologies. These simulations were then scaled to more than 1000 cosmological parameter sets and 10 redshifts. We then train a neural network that quickly predicts the nonlinear power spectrum. We estimate that our emulator reaches a precision of about $5-10\%$ to $k=10 \, \ihMpc$. We refer to Zennaro et al. (in prep.) for all the details.


We now check the differences in the cosmology inference using either {\code HALOFIT} or {\code BACCOemu}, when modelling the DES Y3 shear correlation functions down to angles $\theta = 2.5 \, {\rm arcmins}$. For this test, we use an NLA model and do not include baryonic processes to isolate the effects of the non-linear matter power spectrum. Fig.~\ref{fig:halofit} shows how the $1\sigma$ and $2\sigma$ credible levels of $S_8$ obtained with {\code BACCOemu} are systematically higher with respect to {\code HALOFIT} by almost $0.5\sigma$. When applying DES scale cuts, the difference is slightly smaller, about $0.4\sigma$, due to a smaller amount of non-linearities in the data vector. Thus, we find a larger bias given by non-linearities with respect to \cite{Krause2021}, arguably because of the larger cosmological parameters space used here, where the accuracy of {\code HALOFIT} is lower.  

\section{Comparison against hydrodynamical simulations}
\label{app:hydro_compa}

\begin{figure}
\includegraphics[width=0.4\textwidth]{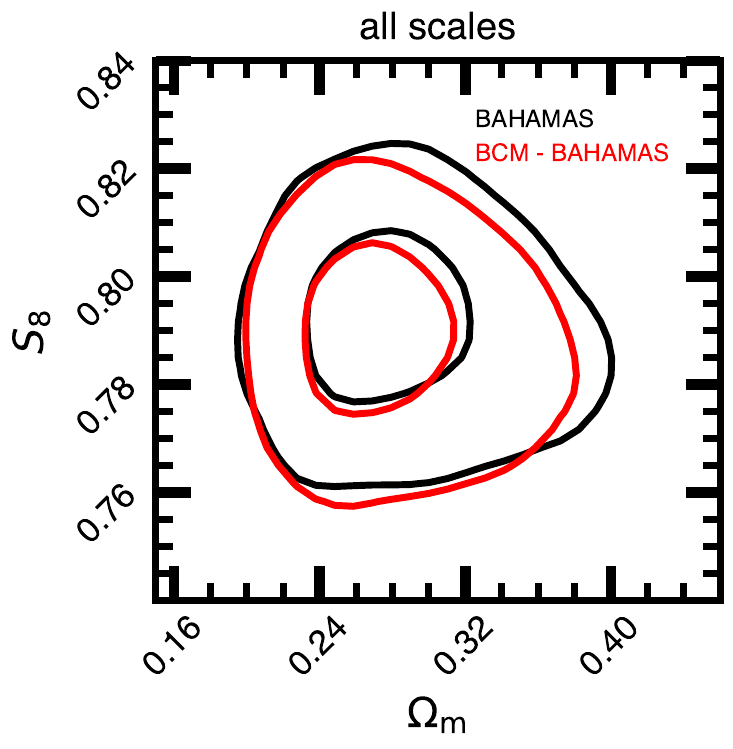}
\caption{$1\sigma$ and $2\sigma$ credible levels on $S_8$ and $\Omega_{\rm m}$ obtained by assuming the baryonic feedback measured in BAHAMAS (black), and using the BCM with the best-fitting parameters to the matter power spectrum at $z=0$ of BAHAMAS (red).}
\label{fig:fixed_bar}
\end{figure}

The baryonification emulator in {\code BACCOemu} has been extensively tested against several hydrodynamical simulations \citep{Arico2020c,Chen2023}. 
Here, we further test the ability of the emulator to mimic the impact of the baryonic processes as predicted by a complex hydrodynamical simulation. 
To do so, we take the suppression in the matter power spectrum measured in BAHAMAS \citep{McCarthy2017,McCarthy2018}, at $z \in [0.,3.]$ and $k \in [10^{-2}, 5 \times 10^{2}] \ihMpc$, and use it as a fiducial baryonic suppression for the cosmology inference in the analysis of the cosmic shear measured by DES Y3. 
We compare it to the results that we obtain when employing the BCM, fixing the free parameters to the best-fitting values to the matter power spectrum of BAHAMAS at $z=0$. The best-fitting parameters, obtained by \cite{Arico2020c}, are $\log{M_{\rm c}}=13.62 [\log (\Msun)]$, $\log{\eta_{b}}=-0.33$, $\log{\beta_{b}}=-0.28$, $\log M_{z0,\rm cen}=10.21 [\log (\Msun)]$, $\log{\theta_{\rm inn}}=-0.62$, $\log{\theta_{\rm out}}=0.12$, $\log{M_{\rm inn}}=9.95 [\log (\Msun)]$. 

We show the comparison between the two posteriors in the $S_8$-$\Omega_{\rm m}$ plane in Fig.~\ref{fig:fixed_bar}. The cosmological constraints are in very good agreement, highlighting the remarkable flexibility and accuracy of the BCM emulator, and assessing its adequacy to study DES Y3 data. 
In particular, this shows the negligible impact of the relatively restricted regime of validity of the emulator ($z \in [0.,2.5]$ and $k \in [10^{-2}, 10\,\ihMpc$]), and of the assumption of no redshift evolution of the baryonification parameters\footnote{The parameters do not change with redshift, but their effects on the power spectrum vary with redshift due to e.g. the halo mass function evolution.}.

\section{Cosmological parameters}
\label{app:cosmo_posteriors}

\begin{figure*}
\includegraphics[width=0.9\textwidth]{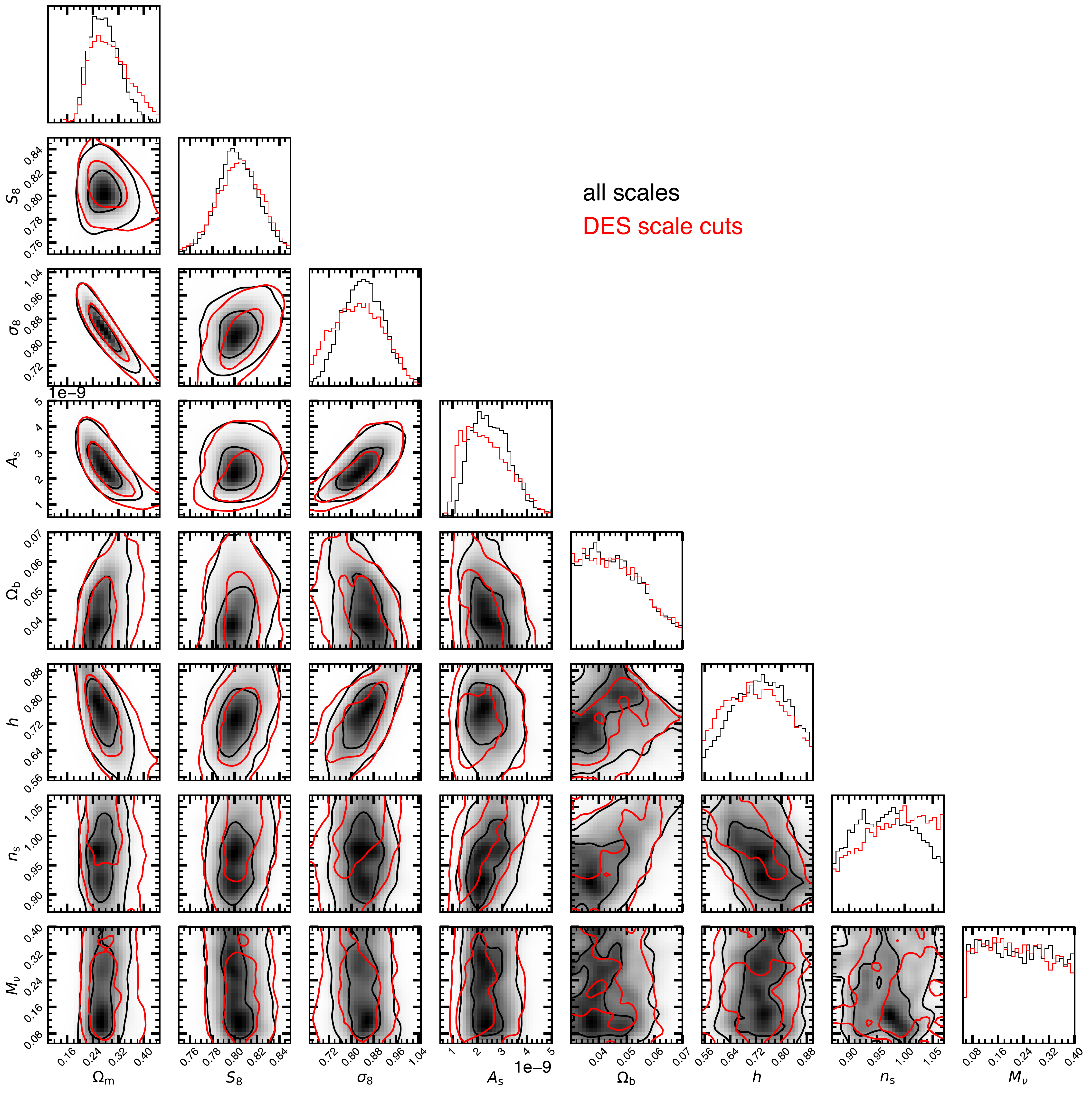}
\caption{Posteriors of the cosmological parameters, obtained analysing all the angular scales of DES Y3 cosmic shear with our fiducial model (black). For comparison, we over plot the $1\sigma$ and $2\sigma$ credible levels obtained with DES scale cuts (red).}
\label{fig:cosmo_pars}
\end{figure*}

We present in Fig.~\ref{fig:cosmo_pars} the posteriors of all the free cosmological parameters in our analyses, when applying DES scale cuts and when analysing all the scales. The main gain in constraining power when analysing the small scales is in 
$\Omega_{\rm m}$, and $S_8$ or alternatively $\sigma_8$. We get somewhat better constraints also in the Hubble parameter. Specifically, when analysing the full data vector, we constrain the normalisation of the primordial linear matter power spectrum $A_{\rm s} = \left( 23.1^{+8.9}_{-7.0} \right) \times 10^{-10}$ and the amplitude of the linear matter density fluctuations $\sigma_8=0.843^{+0.072}_{-0.081}$. 
We also weakly constrain the dimensionless Hubble constant $h=0.742^{+0.096}_{-0.098}$, and the tilt of the primordial matter power spectrum, $n_{\rm s}=0.971^{+0.057}_{-0.071}$. The baryonic density remains unconstrained (although high values of $\Omega_{\rm b}$ are slightly disfavoured), as well as the sum of neutrino masses $M_{\nu}$. \\

\begin{table*}
    \begin{tabular}{c c c c c}
    Model & \multicolumn{2}{c}{$S_8$} & \multicolumn{2}{c}{$\Omega_{\rm m}$}\\ \hline

    This work & DES scale cuts & all scales & DES scale cuts & all scales \\ \hline
    NLA BCM7 (fiducial) & $0.804^{+0.021}_{-0.021}$ & $0.802^{+0.020}_{-0.018}$ & $0.291^{+0.074}_{-0.055}$ & $0.275^{+0.053}_{-0.045}$\\
    NLA GrO & $0.800^{+0.019}_{-0.021}$ & $0.787^{+0.014}_{-0.015}$ & $0.287^{+0.064}_{-0.052}$ & $0.270^{+0.063}_{-0.040}$\\
    NLA BCM1 & $0.803^{+0.019}_{-0.019}$ & $0.798^{+0.016}_{-0.018}$ & $0.285^{+0.067}_{-0.053}$ & $0.274^{+0.060}_{-0.045}$\\
    NLA BCM-extreme & $0.841^{+0.025}_{-0.027}$ & $0.829^{+0.023}_{-0.022}$ & $0.286^{+0.069}_{-0.053}$ & $0.300^{+0.060}_{-0.054}$\\ \hline
    
    TATT BCM7 & $0.792^{+0.024}_{-0.026}$ & $0.781^{+0.028}_{-0.037}$ & $0.292^{+0.068}_{-0.052}$ & $0.249^{+0.057}_{-0.039}$\\
    TATT GrO & $0.788^{+0.024}_{-0.027}$ & $0.756^{+0.025}_{-0.027}$ & $0.277^{+0.062}_{-0.047}$ & $0.270^{+0.061}_{-0.042}$\\
    TATT BCM1 & $0.791^{+0.024}_{-0.029}$ & $0.780^{+0.025}_{-0.037}$ & $0.276^{+0.068}_{-0.050}$ & $0.250^{+0.050}_{-0.035}$\\
    TATT BCM-extreme & $0.825^{+0.030}_{-0.031}$ & $0.815^{+0.029}_{-0.037}$ & $0.283^{+0.072}_{-0.054}$ & $0.258^{+0.058}_{-0.039}$\\ \hline
    DES Collaboration &  &  &  &  \\ \hline
    DES NLA & $0.784^{+0.022}_{-0.023}$ & - & $0.310^{+0.069}_{-0.057}$ & -\\
    DES NLA + SR & $0.773^{+0.020}_{-0.020}$ & - & $0.289^{+0.054}_{-0.046}$ & -\\
    DES TATT & $0.759^{+0.032}_{-0.038}$ & - & $0.293^{+0.064}_{-0.052}$ & -\\
    DES TATT + SR (fiducial) & $0.759^{+0.023}_{-0.024}$ & - & $0.285^{+0.058}_{-0.048}$ & -\\ \hline 
    \end{tabular}
     \centering
    \caption{Constraints (median of the 1D marginalised posteriors and respective 34th percentiles) on $S_8$ and $\Omega_{\rm m}$ obtained applying DES scale cuts and using all the angular scales. We explore different models, with NLA or TATT intrinsic alignment, gravity-only, or applying a baryonification with 1 or 7 free parameters (BCM1, BCM7, respectively). We also include a model with an extreme implementation of AGN feedback. We report for comparison the constraints we get analysing the DES Collaboration chains, with and without shear ratios (SR). }
\label{tab:cum_params}
\end{table*}

We also report here Tab.~\ref{tab:cum_params}, where we quote the median of the 1D marginalised posteriors on $S_8$ and $\Omega_{\rm m}$ and their respective 34th percentiles (as opposed to the maximum posterior of Tab.~\ref{tab:model_params}). We report for comparison the values we get when analysing the DES Collaboration chains \citep{Secco2022,Amon2022}, as in the official DES papers they typically report the mean values instead. 

\section{Baryonic parameters}
\label{app:bcm_posteriors}

\begin{figure*}
\includegraphics[width=0.9\textwidth]{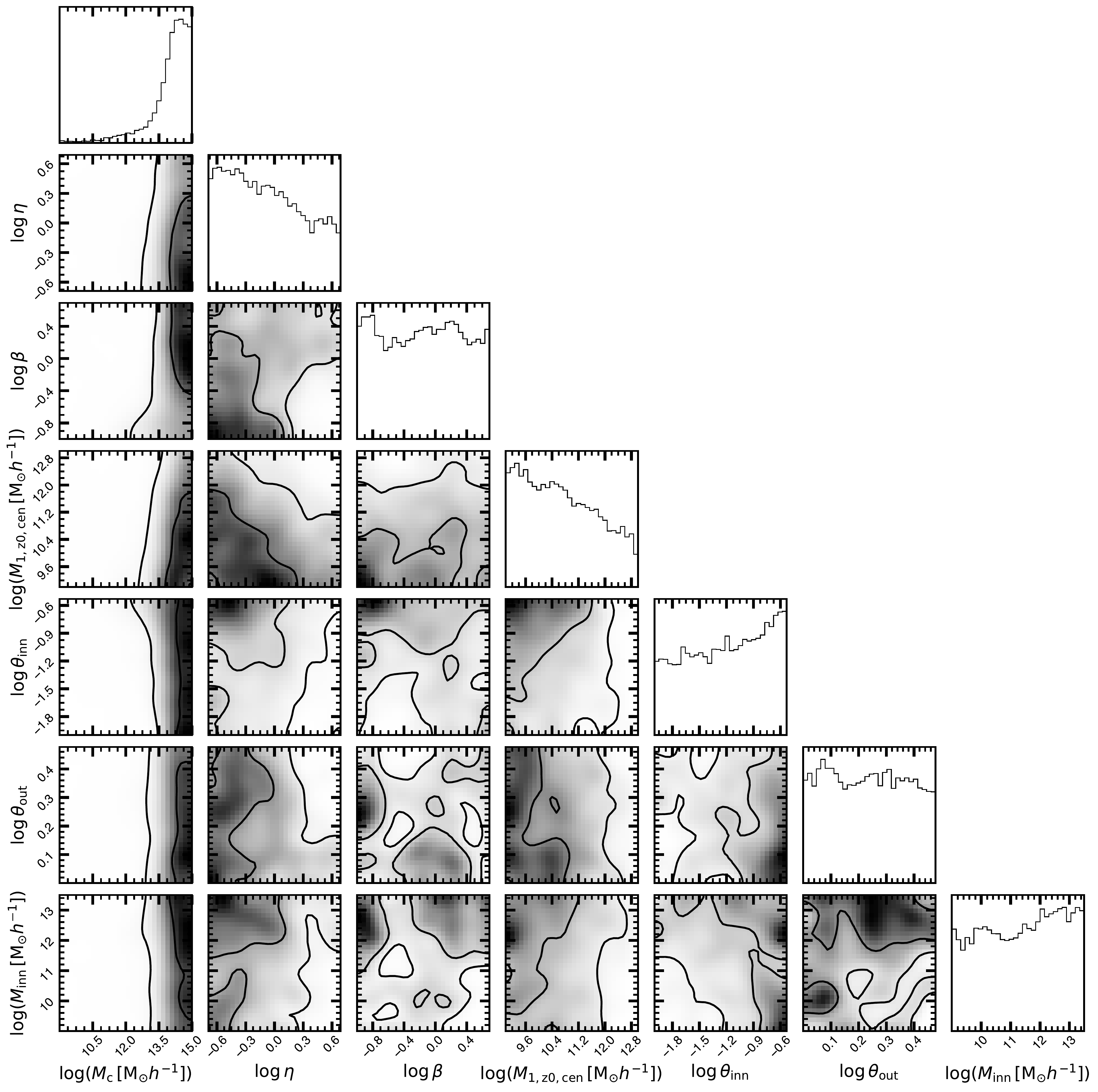}
\caption{Posteriors of the baryonic parameters, obtained analysing all the angular scales of DES Y3 cosmic shear with our fiducial model.}
\label{fig:bcm_pars}
\end{figure*}

In the main text we have reported the posterior of $M_{\rm c}$, the only baryonic parameter strongly constrained by the cosmic shear of DES Y3. Here we report for completeness the posteriors of all the 7 baryonic parameters available in {\code BACCOemu}\footnote{All the baryonic parameters are sampled in log-space.}.

The parameters are: $\eta$, which regulate the baryonic feedback range; $\beta$, the slope of the gas fraction - halo mass; the gas density shape ($\theta_{\rm inn}$,$\theta_{\rm out}$,$M_{\rm inn}$); and $M_{z0,\rm cen}$, the galaxy-halo mass relation. We show in Fig.~\ref{fig:bcm_pars} their posteriors. $M_{\rm c}$ is the only parameter significantly constrained: we obtained $\log M_{\rm c}=14.38^{+0.60}_{-0.56} \, [\log (\Msun)]$. The constraints are robust to the intrinsic alignment model assumed.

\section{Tidal-Alignment \& Tidal-Torque}
\label{app:ia_info}

\begin{figure*}
\includegraphics[width=0.5\textwidth]{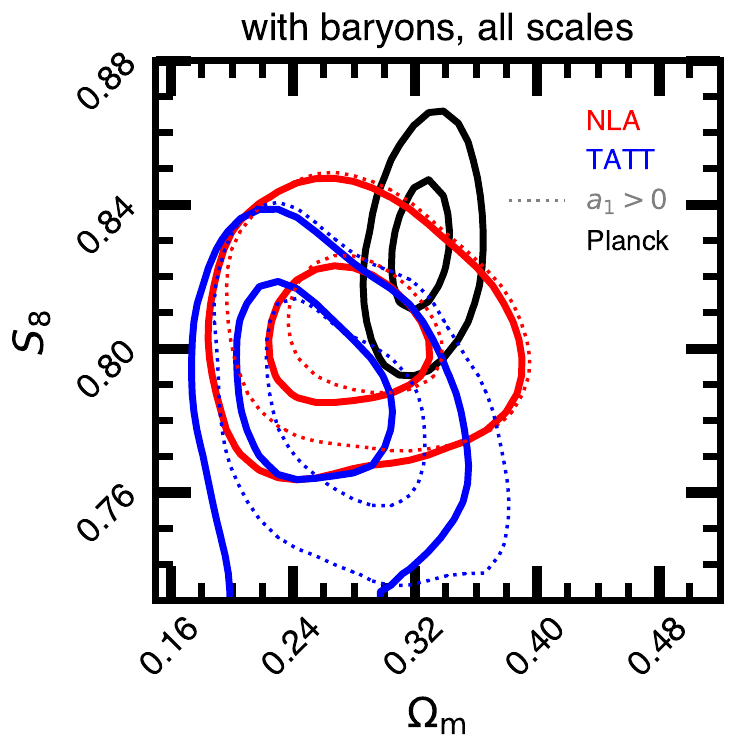}
\caption{Posteriors of $S_8$ of DES Y3 cosmic shear, when employing as intrinsic alignment model NLA (red) or TATT (blue), compared to Planck TT+TE+EE+lowE. To model DES Y3 cosmic shear we have considered a 7-free parameters baryonification. We show with dotted lines the posteriors that we obtain when imposing a physically-motivated prior on the intrinsic alignment amplitude $a_1 \in [0,5]$.}
\label{fig:planck_tatt}
\end{figure*}

\begin{figure*}
\includegraphics[width=0.9\textwidth]{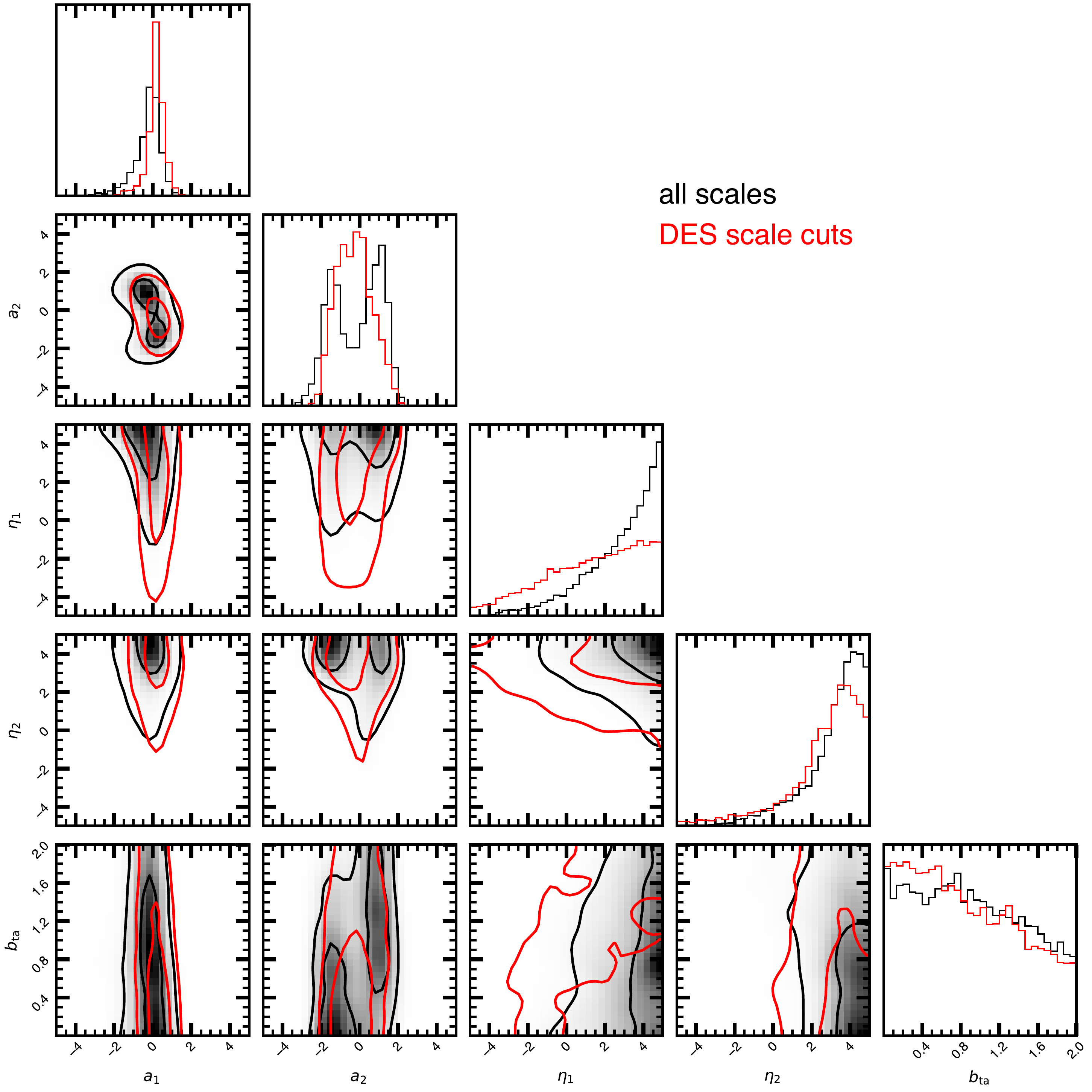}
\caption{Posteriors of TATT parameters when analysing all the angular scales of DES Y3 cosmic shear with our fiducial model (black). For comparison, we overplot the $1\sigma$ and $2\sigma$ credible levels obtained with DES scale cuts (red).}
\label{fig:ia_pars}
\end{figure*}

The Tidal-Alignment \& Tidal-Torque (TATT) model is inspired by perturbation theory and aims to predict the intrinsic alignment of galaxies by considering a perturbative expansion of the cosmic density and tidal fields \citep{Blazek2019}. There are three terms considered: a linear term (tidal alignment), a quadratic term (tidal torque), and a cross term, with a total of 5 free parameters $a_1$, $\eta_1$, $a_2$, $\eta_2$, $b_{\rm TA}$. We have reported the respective equations in \S\ref{sec:ia_define}. 

It is not clear on which scales the NLA and TATT are valid, since the theoretical limits given by linear and perturbation theory are practically overcome by using a full non-linear power spectrum. 
Generally, this is dependent on the galaxy sample analysed, since red galaxies are expected to have higher tidal alignment amplitudes ( $a_1 \in \left[ 3, 5 \right]$) than blue galaxies ($a_1 \in \left[ 0,1 \right]$ ) \citep[][]{Samuroff2019,Samuroff2022}.

The cosmic shear analysis made by the DES Collaboration on year 3 data has highlighted a low amplitude of intrinsic alignment $a_1=-0.47^{+0.30}_{-0.52}$ and $a_2=1.02^{+1.61}_{-0.55}$, much lower than the cosmological signal \citep{Secco2022,Amon2022}. 
Indeed, a model without an intrinsic alignment is marginally preferred with respect to NLA and TATT by Bayesian evidence. Therefore, \cite{Secco2022,Amon2022} conclude a posteriori that both NLA and TATT are suited for the analysis of DES Y3 data. 

Here we are employing smaller scales than in the analysis of \cite{Secco2022,Amon2022}, thus we reassess the impact of scale cuts on the intrinsic alignment parameters after marginalising over 7 baryonic parameters. We have checked that the marginalisation only slightly degrades the TATT parameters constraints, without shifting them in any particular direction. 

We show the posteriors of the TATT parameters in Fig.~\ref{fig:ia_pars}, with and without DES scale cuts (black and red lines, respectively). With scale cuts, we obtain $a_1=0.13^{+0.46}_{-0.26}$ and $a_2=0.00^{+0.48}_{-1.42}$, whereas we obtain $a_1=-0.01^{+0.49}_{-0.63}$ and a fully bimodal $a_2$ when using all the scales. The amplitudes we find are consistent with zero in both cases. As noticed in \cite{Secco2022,Amon2022}, $a_2$ presents a multi-modal posterior. In contrast to \cite{Secco2022,Amon2022}, which reports that adding more data points at small scales (their \say{$\Lambda$CDM-optimised} case) the multimodality was alleviated, we find that adding small scales the multi-modality is exacerbated. According to \cite{Amon2022}, this multi-modality is given by internal degeneracies in the TATT model and it can degrade and shift the posterior on $S_8$. 

In Fig.~\ref{fig:planck_tatt} we show the posterior projected onto the $S_8$ - $\Omega_{\rm m}$ plane when assuming NLA and TATT, and comparing it to Planck. Employing TATT we get $S_8=0.788^{+0.027}_{-0.035}$, a looser constraint with respect to NLA. We see that the posterior shifts also lower $S_8$ and $\Omega_{\rm m}$, at $\approx 2 \sigma$ from Planck. 
This could be an effect of the degeneracy between $a_1$ and $a_2$ discussed above. 
As noted by \cite{Secco2022,Amon2022}, one would expect the intrinsic alignment amplitude $a_1 \ge 0$ in absence of systematics. Therefore, we impose $a_1 \ge 0$ as a prior, and obtain the dotted contours in Fig.~\ref{fig:planck_tatt}. With NLA, this new prior slightly shifts ($0.1\sigma$) the $S_8$ contours towards Planck, whereas with TATT the largest effect is a shift of $0.4\%$ in $\Omega_{\rm m}$.

\section{Shear bias \& Photo-z uncertainties}
\label{sec:deltaz_posteriors}

\begin{figure*}
\includegraphics[width=0.9\textwidth]{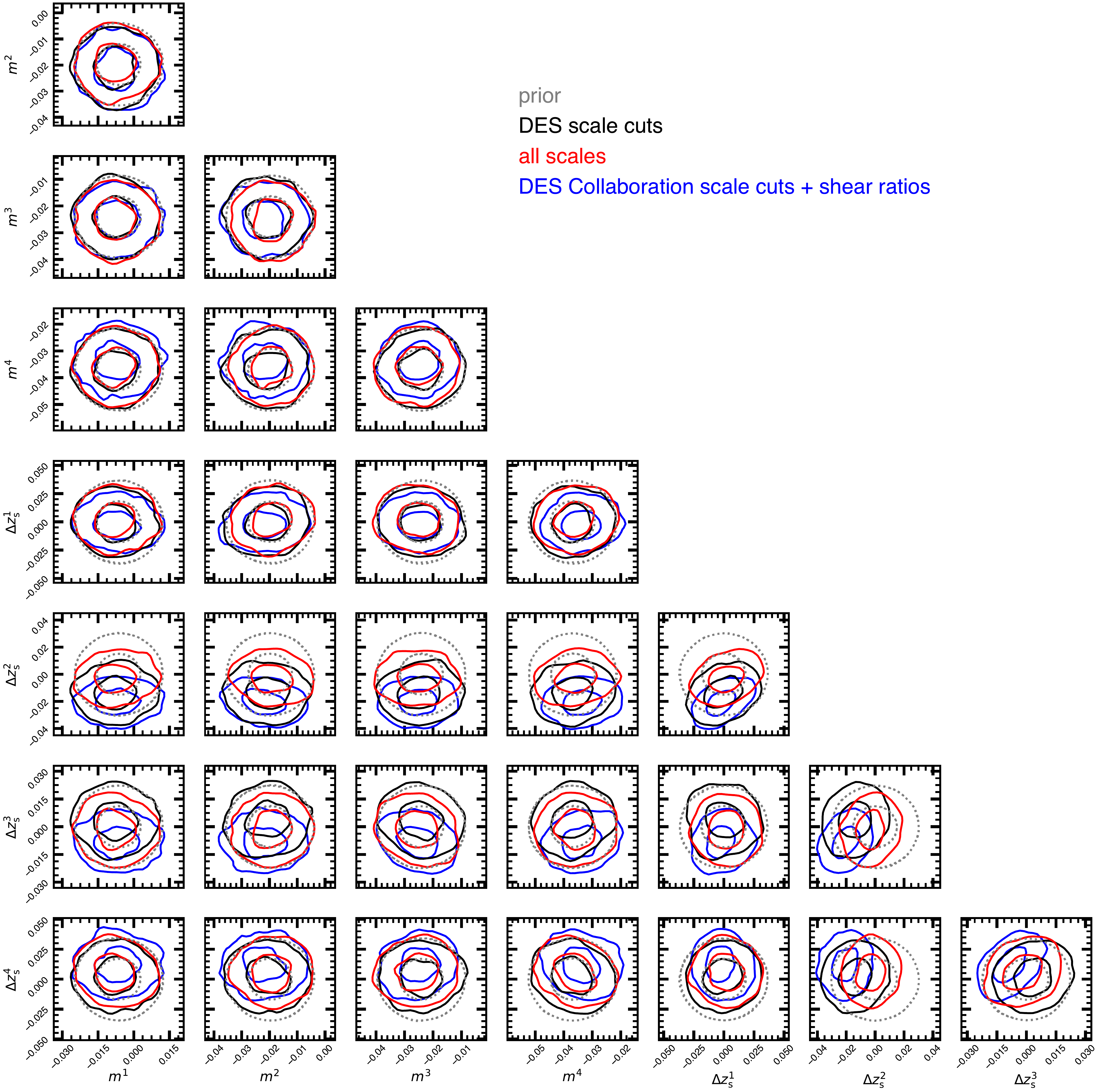}
\caption{Priors (gray dotted lines) and posteriors of the shear biases $m^{\rm i}$ and photo-z uncertainties $\Delta z_{\rm s}^i$, obtained with our fiducial for baryons and intrinsic alignment (BCM \& NLA, respectively). We display the cases when applying DES scale cuts (black) and using all angular scales (red), and compare against the results obtained by the DES Collaboration using NLA with the addition of the shear ratios likelihood \citep[blue, ][]{Secco2022,Amon2022}.} 
\label{fig:deltaz_pars}
\end{figure*}

We display in Fig.~\ref{fig:deltaz_pars} the parameter posteriors of the shear biases $m^{\rm i}$, that take accounts of galaxy shapes miscalibration and blending, and the photo-z uncertainties $\Delta z_{\rm s}^i$, that is the shift of the source redshift distribution in the $i$ bin. 
We show the posteriors obtained with and without scale cuts, using NLA and marginalising over 7 free baryonic parameters. 
We compare against the respective informative Gaussian priors, and additionally to the posteriors obtained by the DES Collaboration adding the likelihood of the lensing shear ratios. Such ratios are shown to be particularly sensitive to the redshift distribution of the source galaxies \citep{Sanchez2022,Amon2022}.

The posteriors of the shear biases are all in good agreement, in some cases, e.g. $m^4$, analysing the small scales gives an improvement in the constraints similar to the addition of the shear ratios. 

Also the $\Delta z_{\rm s}^i$ posteriors are in good agreement with the priors. We note how the constraints obtained with shear ratios are generally tighter with respect to the ones with scale cuts, and in agreement with our full-scale analysis. In some cases, notably $\Delta z_{\rm s}^2$ and $\Delta z_{\rm s}^3$, the results we obtain analysing the full data vector and the results obtained by the DES Collaboration using large scales and shear ratios are slightly in tension, being our constraints more in agreement with the priors. 

\end{appendix}
\label{lastpage}
\end{document}